\documentclass[aps,reprint,amsmath,amssymb]{revtex4-1}

\newcommand{\unit}[1]{\ensuremath{\, \mathrm{#1}}}
\newcommand{\etal}{\textit{et al. }}
\newcommand{\ie}{\textit{i}.\textit{e}.}
\newcommand{\eg}{\textit{e}.\textit{g}.}

\usepackage{graphicx}
\usepackage{dcolumn}
\usepackage{bm}

\begin{document}

\preprint{}

\title{Thermal Fluctuations in Nuclear Pasta}

\author{M. E. Caplan} \email{mecapl1@ilstu.edu}
\author{C. R. Forsman}\affiliation{
Illinois State University, Department of Physics, Normal IL 61790 USA}%
\author{A. S. Schneider}%
\affiliation{Department of Astronomy and the Oskar Klein Centre, Stockholm 
University, AlbaNova, SE-106 91 Stockholm, Sweden}


\date{\today}

\begin{abstract}

Despite their astrophysical relevance, nuclear pasta phases are relatively unstudied at high temperatures. We present molecular dynamics simulations of symmetric nuclear matter with several topologies of `lasagna' at a range of temperatures to study the pasta-uniform transition. 
Using the Minkowski functionals we quantify trends in the occupied volume, surface area, mean breadth, and Euler characteristic. The amplitude of surface displacements of the pasta increase with temperature which produce short lived topological defects such as holes and filaments near melting, resulting in power laws for increasing surface curvature with temperature. We calculate the static structure factor and report the shear viscosity and thermal conductivity of pasta, finding that the shear viscosity is minimized at the melting temperature. These results may have implications for the thermoelastic properties of nuclear pasta and finite temperature corrections to the equation of state at pasta densities.

\end{abstract}

\maketitle

\section{Introduction}\label{sec:int}

As matter is compressed and the density approaches nuclear saturation, it is energetically favorable for nucleons to rearrange from spheres into more complicated shapes such as cylinders and sheets which may contain millions of nucleons, resembling spaghetti and lasagna. These nuclear pasta phases exists on the QCD phase diagram as a transition between isolated nuclei and uniform matter at relatively low temperatures ($T\lesssim \mathcal{O}(15$ MeV)) \cite{schuetrumpf2013time,astromaterials,PhysRevC.100.025803}. 

Work studying the behavior of nuclear pasta at finite temperature is well motivated observationally, as the inner crusts of neutron stars may form nuclear pasta in many astrophysically relevant scenarios. If present, pasta may affect many transport properties and astrophysical observables. To name a few: the electron transport and conductivities in pasta may impact magnetic field evolution and thermal evolution \cite{pons2013highly,PhysRevLett.114.031102}, the elastic properties of pasta may set the maximum mass quadrupole which can be a continuous source of gravitational waves \cite{CaplanPRL,abbott2019searches,pethick2019dense}, and dark matter annihilation in the pasta layer has even recently been proposed as a detectable heat source \cite{acevedo2019cooking}.  

Properties of pasta at finite temperatures and transport properties near the melting temperature may be relevant to the evolution of remnants in neutron star mergers. Recent numerical simulations by Hanauske \etal\ predict nuclear matter at pasta densities to be present approximately 10 to 14 km from the center of the merger remnant with temperatures between 10 and 20 MeV for tens of milliseconds postmerger \cite{Hanauske:2019qgs}. As time-dependent Hartree-Fock simulations by Schuetrumpf \etal\ predict a melting temperature between 10 and 14 MeV, one may expect melted or partially melted crusts in merger remnants \cite{PhysRevC.90.055802,PhysRevC.95.055804}. Any temperature dependence in the transport properties, especially near the melting temperature, may impact the post-merger gravitational ringdown. This motivates the study of nuclear pasta phases near the melting temperature and the calculation of transport properties which may be of interest to numerical simulations of mergers. 

While the exact geometry of pasta phases is likely a sub-dominant contributor to the heat capacity and thermal conductivity of nuclear matter, thermal fluctuations in pasta near the melting temperature may produce long range disorder which may affect other transport properties.
Finite temperature defects and thermal excitation of phonons, disrupting long range order in pasta, will effect the static and dynamic response factors $S(\bf{q})$ and $S(\bf{q},\omega)$ and has been studied in a few specific cases by Schneider \etal\ \cite{PhysRevC.93.065806} and Horowitz \etal\ \cite{PhysRevLett.114.031102}. 
As an illustrative example of the kinds of defects one might expect, consider analogs from terrestrial physics. Pasta resembles block copolymers, which are known to have complex geometric phases including defects \cite{van2009thin}.
Distortions of the pasta surface may include topological defects such as filaments or holes \cite{CaplanPRL}.  Filaments or holes disrupt local order similar to interstitials, vacancies, and impurity substitutions in conventional crystal lattices. Helicoids which connect lasagna sheets, directly analogous to screw dislocations in both liquid crystals and conventional crystal lattices, are now well studied in pasta MD and are also resolved in analog terrestrial experiments with biological membranes \cite{PhysRevC.94.055801,PhysRevLett.113.188101,PhysRevLett.114.031102}. Larger scale dislocations such as stacking faults may also be present at domain boundaries in `polycrystalline' pasta, which may be expected at the mesoscale \cite{Schneider:18,CaplanPRL}. Some low-angle or tilt boundary defects between domains have been resolved in MD simulations by Caplan \etal\ \cite{CaplanPRL} and Schneider \etal\ \cite{PhysRevC.93.065806}. 

We report on simulations of nuclear pasta in symmetric nuclear matter $(Y_e = 0.5)$ in this work, which has not yet been well characterized in our model \cite{PhysRevC.90.055805,PhysRevC.88.065807}. While the electron (proton) fraction in neutron stars may be $Y_e \sim 0.1$ or less, matter may reach pasta densities with relatively high proton fractions in a supernova which could have important consequences for neutrino trapping and the evolution of the proto-neutron star \cite{Horowitz2016supernova}. Many pasta models predict the lasagna/slab phase will form even at much lower proton fractions, so even if the quantitative results we obtain do not match the low proton fraction lasagna, the qualitative results we obtain may extend to much lower proton fractions \cite{Grill2012}. If the exact thermodynamic conditions simulated in this work are not found in any astrophysical environment, these results may still be useful to future authors modeling transport properties of pasta as limiting cases of high temperature and high proton fraction (\eg\ for corrections to the nuclear surface energy in pasta for supernova codes).

In this work we study the pasta phases near the melting temperature with molecular dynamics simulations. Sec. \ref{sec:sim} describes our model formalism, Sec. \ref{sec:res} presents our simulations. In Sec. \ref{ssec:sq} we present calculations of the static structure factor which we use to compute observables in Sec. \ref{ssec:obs}. Sec. \ref{sec:dis} summarizes. 




\section{Model and Formalism}\label{sec:sim}

\subsection{Semi-classical Pasta Model}

The nuclear pasta model used in this work is the same as in a large body of past work, and is discussed in detail in refs. \cite{PhysRevC.69.045804,PhysRevC.88.065807,PhysRevC.91.065802,astromaterials}. We briefly review it here for completeness. 

We simulate using the Indiana University Molecular Dynamics (IUMD) code, version 6.3.1, a CUDA-Fortran code which runs on the Big Red II supercomputer at Indiana University. The semi-classical model treats nucleons $i$ and $j$ as point particles (with periodic separation $r$) which interact via the two-body potential

\begin{equation}\label{eq:pot}
 V_{ij}(r)=ae^{-r^2/\Lambda}+[b \pm c]e^{-r^2/2\Lambda}+\frac{e_i e_j }{r}e^{-r/\lambda}\,.
\end{equation}

\noindent The parameters $a$, $b$, $c$, and $\Lambda$ are given in Tab. \ref{tab:param} and were chosen by Horowitz \etal\ to reproduce known properties of nuclear matter near saturation, while $\lambda$ is the Coulomb screening length due to the electron gas (included in our simulations only through this term) and is fixed at 10 fm as in past work \cite{PhysRevC.69.045804}. 

\begin{table}[h]\label{tab:param}
\caption{Model parameters for Eq. \ref{eq:pot}}
 \begin{ruledtabular}
 \begin{tabular}{cccc}
   $a$     &  $b$    &  $c$   & $\Lambda$ \\
   \hline
   110 MeV & -26 MeV & 24 MeV & 1.25 fm$^2$
 \label{Table1}
 \end{tabular}
\end{ruledtabular}
\end{table}

These potentials are qualitatively similar to a binary Lennard-Jones mixture, as the $b+c$ ($b-c$) term sets a weak (strong) attraction between like (unlike) nucleons. The final term is a long range screened Coulomb repulsion between protons due to their electric charges $e_i$ and $e_j$ ($e_i e_j \approx 1.44$ MeV fm). All simulations in this work use periodic boundary conditions. 


\subsection{Pasta Configurations}\label{ss:pastaconfig}

\begin{figure}[t!]
\centering
\includegraphics[trim=0 45 0 45,clip,width=0.49\textwidth]{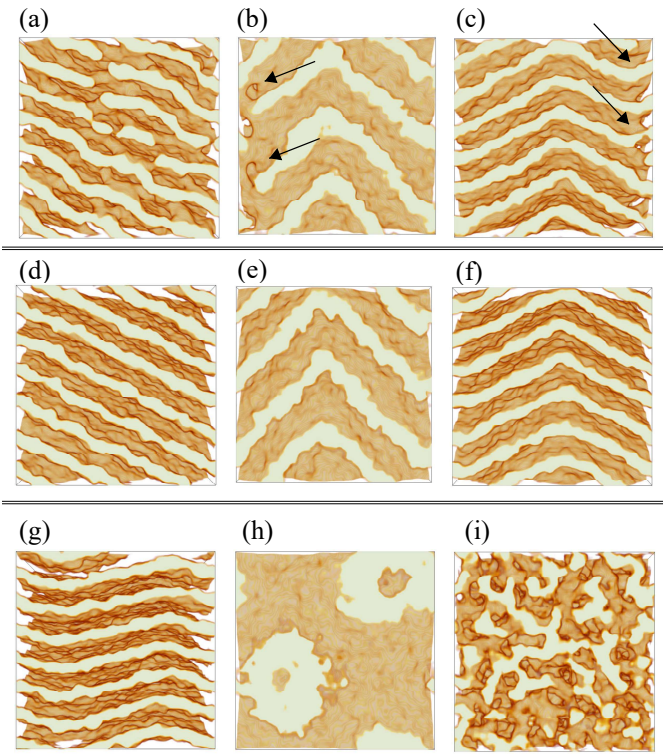}
\caption{\label{fig:init} (Color online) Faces of our pasta configurations at the lowest temperature studied in this work, $T/T_m=0.70$, effectively our initial conditions. The golden surfaces are isosurfaces of charge density of $n_{ch} = 0.03 \unit{fm}^{-3}$ (\ie\ surfaces bounding the region where the protons are most abundant) while the cream fill shows where charge density of $n_{ch} > 0.03 \unit{fm}^{-3}$ within the pasta structure, visible due to the intersection of the pasta structure with the periodic box boundary.  In (a), (b), and (c) show three orthogonal faces of the simulation with `defects.' The defects consist of a wall of helicoids, seen in (a), connecting the plates; in (b) one can see the axis of the helicoids on the left, highlighted by the arrows. In (c), one can see how the following the helicoids cause the plates to bend up at the periodic boundary on the right, connecting the seemingly distinct plates. These can be compared with the related configuration without defects, which we call  `nonparallel,' shown in (d), (e), and (f). In (g) and (h) we show two views of the configuration which is `parallel' with the box, and in (i) we show a configuration above the melting temperature.}	
\end{figure}

We study planar phases of nuclear pasta called lasagna which are equivalent to lamellar phases in block copolymer studies \cite{van2009thin}. Lasgana has an obvious advantage for resolving thermal fluctuations quantitatively. This is the only phase where we should expect both principal curvatures $k_1$ and $k_2$ of the pasta surface to be locally zero everywhere in the ground state. Thus, the topological characterizations we use (which integrate curvature over the surface) will be zero in the ground state. This means we can readily resolve absolute deviations from zero due to thermal fluctuations no matter how small. In contrast, a surface with finite ground state curvature will need thermal fluctuations whose curvature is comparable to the ground state curvature to be easily resolved. 

Our initial conditions are three variations of the planar `lasagna' phase, shown in Fig. \ref{fig:init}. These configurations were used in prior work to study the elastic properties of nuclear pasta \cite{CaplanPRL}. They are (i) a set of plates with a helicoid wall (`defects’), (ii) a set of plates with no defects which are not aligned with the simulation boundary (`nonparallel’), and (iii) a set of that are aligned with the simulation boundary (`parallel’). 
The `defects' simulation with the helicoidal wall was produced from random initial conditions and is fully topologically connected, meaning that there is a path between any point on the surface of the pasta structure to any other.  
The `nonparallel' configuration was produced from a simulation which sheared the `defects' configuration until the helicoids broke, and then contracted back to a cubic box. Through the periodic boundary, there are three topologically distinct plates in the simulation volume. Lastly, the `parallel' simulation in which the plates are aligned with the box boundary was generated by including a sinusoidal external potential when first initialized and contains seven topologically distinct structures. This external sinusoidal potential is not included in any further simulations described in this work and is not required for this structure to remain stable. More detailed information about these configurations is presented in ref. \cite{CaplanPRL}.

All simulations in this work contain 102400 nucleons in a cubic volume at a nucleon density of $n=0.05 \unit{fm}^{-3}$, approximately a third of saturation density where most models predict the existence of the lasagna phase \cite{PhysRevC.88.065807,PhysRevC.95.055804,PhysRevC.90.055802}. In contrast to past work with our model which focused on proton fractions of $Y_P = 0.4$, we report on simulations of symmetric nuclear matter with equal numbers of protons and neutrons ($Y_P =0.5$). To convert our configurations to this higher proton fraction neutrons were chosen at random to be switched for protons. We use the higher proton fraction because we expect the pasta to be stable for a larger range of temperatures, and also to allow for comparison to ref. \cite{dorso2018phase} whose model is similar to our own and has been characterized in these regimes. 

The three configurations we consider are all similar in energy (per nucleon) and are long lived. As in many glassy systems, there may be many local minima separated by large tunneling barriers in the energy landscape, making our pasta structures long lived even if they are not the true ground state. Taken together, these three structures will allow us to characterize the behavior of thermal fluctuations in nuclear pasta with similar topologies. 

\subsection{Melting Temperature}

We perform one simulation for each topology described in Sec. \ref{ss:pastaconfig} to resolve the melting temperature in our model. These simulations begin using three configurations at $T=1.7$ MeV and are heated by rescaling the velocities to a Maxwell Boltzmann distribution $+\Delta T=10^{-4}$ MeV hotter every $10^{3}$ timesteps. The temperature thus increases to a final temperature of $T=1.8$ MeV after the $10^6$ timesteps of the simulation. The melting transition is resolved at $T_m=1.72$ MeV from these simulations. Caloric curves produced from these simulations (omitted for length) show that the energy per nucleon changes discontinuously, consistent with a first order phase transition. Furthermore, above this temperature the nuclear pasta structure appears to dissolve into a disordered set of filaments with little long range order and large fluctuations. This result is consistent with Fig. 6b in ref. \cite{dorso2018phase}.

With the melting temperature known, we prepare three addition configurations above the melting temperature, at $T=1.8$, $1.9$ and $2.0$ MeV. The initial conditions are largely unimportant, as these configurations are disordered and fluid-like. 
These simulations were run for 100,000 timesteps to allow them to equilibrate;  
the energy converged within 1,000 timesteps suggesting that at these high temperatures our model equilibrates quickly. 

\subsection{Simulations of Thermal Fluctuations}

From the initial configurations described above, we perform a set of 21 simulations from which we calculate the Minkowski functionals and static structure factors to study thermal fluctuations in nuclear pasta. These include a simulation of each of our three topologies at $T=1.2$, $1.3$, $1.4$, $1.5$, $1.6$, and $1.7$ MeV, for a total of 18 simulations below the melting temperature, and one simulation each at $T=1.8$, $1.9$, and $2.0$ MeV to study the behavior above the melting temperature (hereafter we refer to these simulations in units of the model melting temperature, $T_m$=1.72 MeV).\footnote{The minimum temperature is constrained by the model; at low $T$ the semi-classical model undergoes a phase transition to a solid, which we do not regard as physically relevant for nuclear physics, though this phase transition and the behavior of the model at low $T$ may be interesting if this model is used to study analagous systems, such as self-assembly in colloidal mixtures \cite{RevModPhys.89.041002,PhysRevC.94.055801}.}  
These simulations are evolved for $10^5$ MD timesteps and configurations are stored every 100 timesteps for a total of $10^3$ snapshots of the configuration. These simulations are run in the microcanonical ensemble and do not include any temperature renormalizations (unlike most past work with our model).
Video renders of these simulations are available in the supplemental materials (SM) (online at \cite{SM}) while select frames from these are shown in Fig. \ref{fig:heat}.


\subsection{Minkowski Functionals}
 
We study thermal fluctuations in our pasta structures using the normalized Minkowski functionals. In summary, the Minkowski functionals quantify the geometry of the pasta surfaces, including surface curvature, topological connectivity, occupied volume, and surface area and so they are useful for characterizing the pasta model. While they may have limited immediate application to astrophysics, it is possible that future authors interested in corrections to the nuclear equation of state at pasta densities may find them useful when building curvature corrections to the surface energy in nuclear equations of state \cite{PhysRevC.88.065807,RevModPhys.89.041002}.   

In three dimensions, the four Minkowski functionals are proportional to the occupied volume $V_\mathrm{occ}$, surface area $A$, mean breadth $B$, and Euler characteristic $\chi$. The volume and surface area are straightforward to understand while the mean breadth and Euler characteristic depend on the principal curvatures $k_1$ and $k_2$ of the pasta surface $\partial K$. The mean breadth is defined by

\begin{equation}
    B = \frac{1}{4\pi}\int_{\partial K} (k_1 + k_2) dA
\end{equation}

\noindent and measures the average curvature of the bounding surfaces $dA$; it is a surface integral over the mean curvature $(k_1 + k_2)$ on domain $\partial K$. The Euler characteristic is similarly defined, 

\begin{equation}\label{eq:CALSFACE}
    \chi = \frac{1}{4\pi}\int_{\partial K} (k_1 k_2) dA
\end{equation}

\noindent and measures the bounding surface curvature as a surface integral over the Gaussian curvature $(k_1 k_2)$. From the definition of the Gaussian curvature this integral is proportional to the total curvature which is a measure of the convexity ($ \chi < 0$), concavity ($ \chi > 0$), or flatness ($ \chi = 0$) of the bounding surface. By the Gauss-Bonnet theorem this also makes $\chi$ a measure of the topology (connectedness) of the surface. Large negative $\chi$ implies a well connected surface with many tunnels, large positive $\chi$ implies many topologically disconnected surfaces, and zero $\chi$ is reserved for planar structures. We normalize by total surface area to $B/A$ and $\chi/A$ using

\begin{equation}
    A=\int_{\partial K} dA.
\end{equation}

While the exact computational details of our algorithm are very extensive and are beyond the scope of this work, they are laid out in detail in Sec. IIb in ref. \cite{PhysRevC.88.065807}. 
Our nucleons are point-like, so finding bounding surfaces is non-trivial. 
To find them we treat protons as a normal distribution ($\sigma = 1.5$ fm) centered on the particle and calculate the `nucleon density' on a 3D grid of `voxels' (\ie\ a 3D dimensional pixel). This is used to produce a discretized binary image of a configuration; if the nucleon density of the voxel is above a threshold of $n_{th} > 0.03 \unit{fm}^{-3}$ it is considered `occupied' while if it is below threshold it is considered `unoccupied.' The binary occupation of each voxel taken together with that of its nearest neighbors contributes can be used to calculate the Minkowski functionals following the algorithm by Lang \etal\ \cite{lang2001analysis}. For example, $V_\mathrm{occ}$ is simple the number of occupied voxels. The $A$ is proportional to the number of unoccupied voxels which are adjacent to occupied voxels. The curvatures $B$ and $\chi$ are more complicated to compute but similarly follow from calculating occupations of all $2\times2\times2$ subvolumes and summing the curvature contributions from each.
It is worth noting that our Minkowski functionals are technically quantized by this formalism, however, they are at such high resolution that they are effectively continuous for our purposes. We emphasize that the choices of $\sigma$ and $n_{th}$, among others, are the result of a thorough analysis by Schneider \etal\ and have been used extensively in a growing body of work \cite{PhysRevC.88.065807}. 

\section{Results}\label{sec:res}

\subsection{Simulations}

\begin{figure}[t!]
\centering
\includegraphics[trim=0 0 0 0,clip,width=0.49\textwidth]{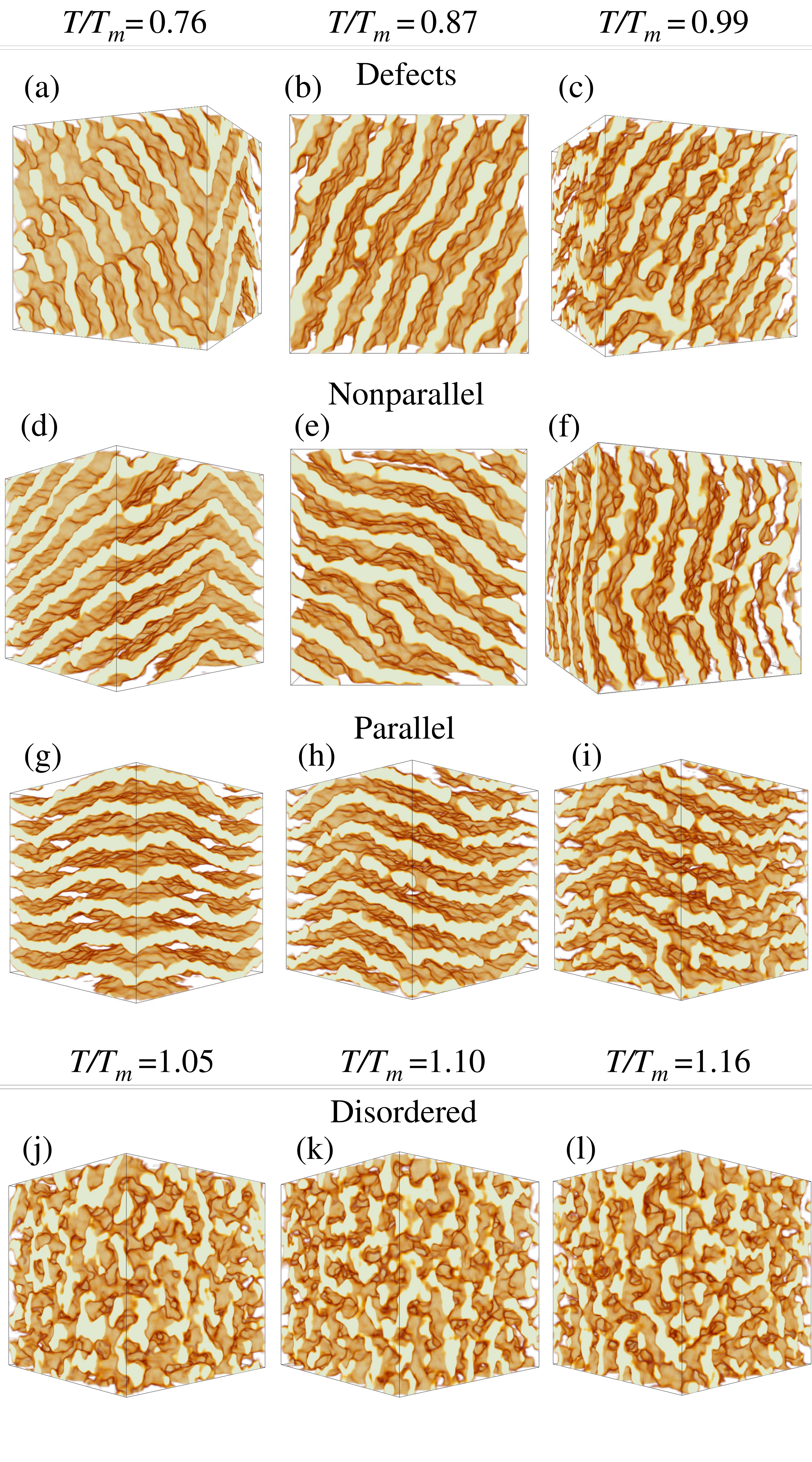}
\caption{\label{fig:heat} (Color online) Faces of our pasta configurations at a range of temperatures. Top three rows show our `defects', `nonparallel,' and `parallel' configurations at three temperatures (columns). The bottom row shows simulations above the melting temperature. See the SM for animations.}
\end{figure}

\begin{figure*}[t!]
\centering
\includegraphics[trim=0 150 150 0,clip,width=0.8\textwidth]{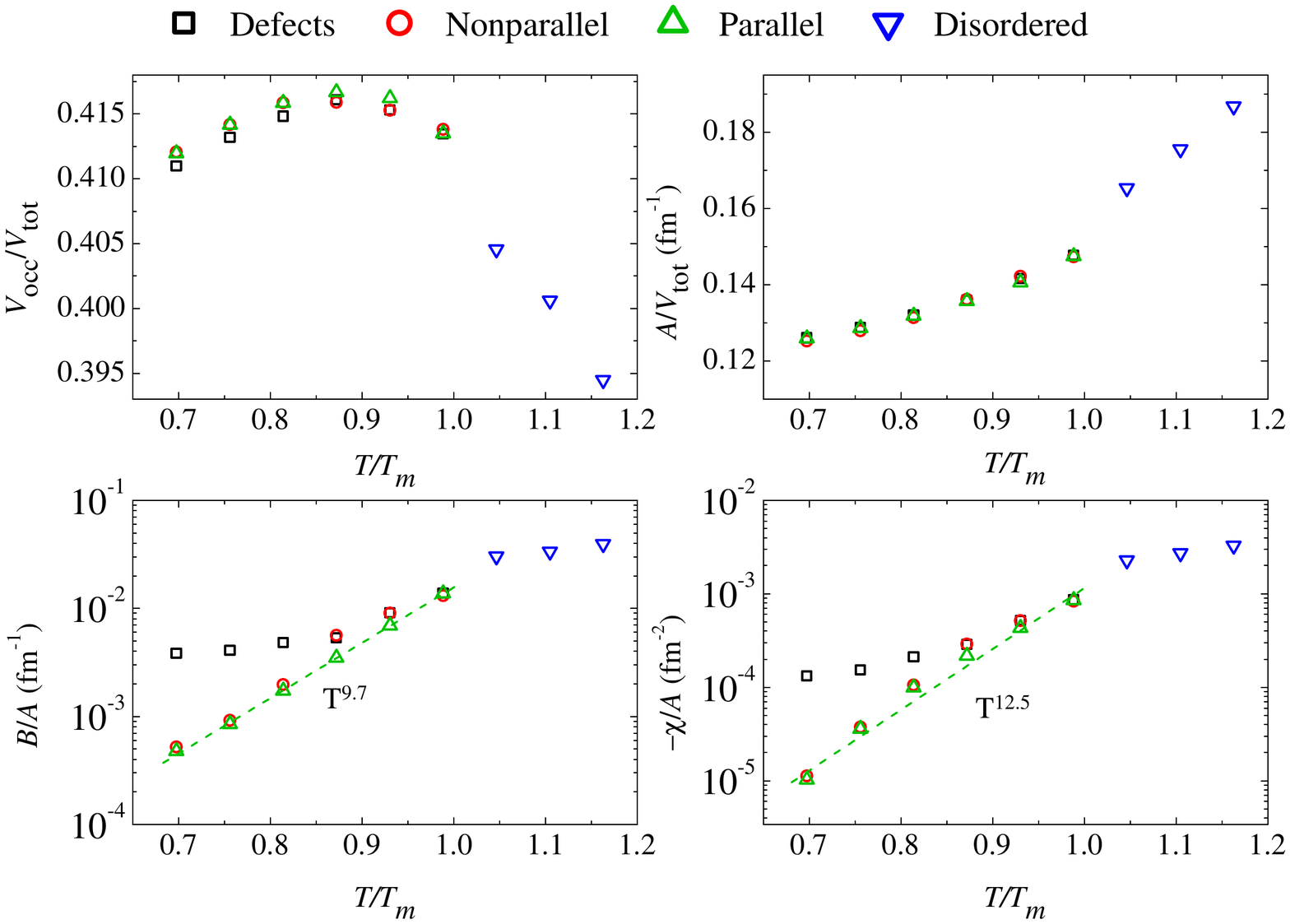}
\caption{\label{fig:min} (Color online) The four normalized Minkowski functionals for our simulations. Clockwise from top right: the occupied volume fraction, surface area density, Euler characteristic, and mean breadth. Note the sign on the units of the Euler characteristic.}	
\end{figure*}

To begin, we describe the qualitative features of the pasta structures in Fig. \ref{fig:heat} and in the SM. At the lowest temperatures considered ($T/T_m=0.70$) all three configurations studied are relatively smooth with little surface roughness or variation, shown in Fig. \ref{fig:init}. Very few holes spontaneously form and their lifetimes are short, appearing in only one or two frames of the simulation before collapsing. They are most easily observed in the nonparallel simulation (center SM). We conclude that the topology is constant and frozen in for configurations below this temperature.
The plate splay is notable as well. The related `defects' and `nonparallel' configurations both show a sharp buckling angle while the plates are nearly planar to either side, while the `parallel' plates show some weak sinusoidal or hyperbolic splay with a length scale of order the box width.

Topological thermal fluctuations become increasingly frequent at higher temperatures ($T/T_m=$0.76 and 0.81), shown in Fig. \ref{fig:heat}(a), (d), and (g). One or a few holes can be seen at almost all times in the SM animation. This is easily seen in both the surface and also in the simulation edges. Discontinuities in the cream surface are due to holes which cross the periodic boundary. Increasing the temperature increases the surface roughness as larger amplitude oscillations become more frequent, though their amplitude does not appear to be sufficiently large to produce filaments connecting the plates with high enough probability to resolve on MD timescales. The magnitude of splay is largely unchanged relative to the lowest temperature considered, though some lateral translation of the plates may have occurred in the `parallel' system. We note that the `defects' appear to migrate in the SM animation at these temperatures. There are two pairs of defects, forming an alternating wall of left handed and right handed defects, visible in Figs. \ref{fig:heat}(a) and \ref{fig:init}(a) and from the top in the LHS of Fig. \ref{fig:init}(b). Past work has argued that these helicoidal ramps tend to experience long range attractive forces, explaining their organization into walls of dipoles of alternating helicity \cite{PhysRevC.94.055801,PhysRevLett.113.188101}. The apparent migration of these helicoids suggests that thermal energy is sufficiently high to overcome the attraction between these ramps and unbind their clustering, but not sufficiently high to dissolve the ramps, which may have implications for the structure of nuclear pasta that forms as neutron star crusts cool and anneal. 
 
Further increasing the temperature ($T/T_m$=0.87), we now resolve the formation and dissolution of filaments which connect the plates, seen in Fig. \ref{fig:heat}(b) and (e). As filaments first appear at higher temperatures than holes, we argue that they experience a higher formation barrier than holes. They have lifetimes comparable to holes or greater. These filamentary fluctuations have significant effects on the topology in all of our simulation. The helicoids dissolve in the `defects' simulations; we observe that the bridges between adjacent plates dissolve over the span of about $10^5$ MD timesteps, while in the `nonparallel' simulation we see the spontaneous formation of helicoidal defects connecting a few plates. These simulation may be near a critical temperature for the formation and dissolution of helicoidal defects.
We also note that the splay of the `nonparallel' configuration has changed, while previously the buckle was sharp in Fig. \ref{fig:heat}(d) it appears more sinusoidal in Fig. \ref{fig:heat}(e), similar to the splay of the `parallel' configuration in Fig. \ref{fig:heat} (g-i). 

Our highest temperature simulations below the melting temperature ($T/T_m=0.93$ and 0.99) show similar behavior for all three configurations, seen in Figs. \ref{fig:heat} (c), (f), and (i). The pasta weakly maintains its coarse long range order as all three configurations show a large number of filaments and holes quickly forming and dissolving. Oscillations in the splay of the plates can be observed on MD timescales, particularly in the `parallel' simulation. 

Above the melting transition ($T/T_m$=1.05, 1.10, and 1.16) the `disordered' simulations all show roughly the same behavior, shown in Figs. \ref{fig:heat} (g-i), having a large number of sponge-like filaments with no long range order or temporal persistence. Notably, the size of filaments in the `disordered' simulations may be smaller at higher temperatures, likely due to a larger number of nucleons entering a gaseous phase between the condensed structures.


\subsection{Minkowski Functionals}

We quantify the evolution in topology using the four Minkowski functionals (normalized by the total volume or surface area where appropriate) in Fig. \ref{fig:min} and interpret each below.

\begin{figure}[t!]
\centering
\includegraphics[trim=100 240 100 235,clip,width=0.38\textwidth]{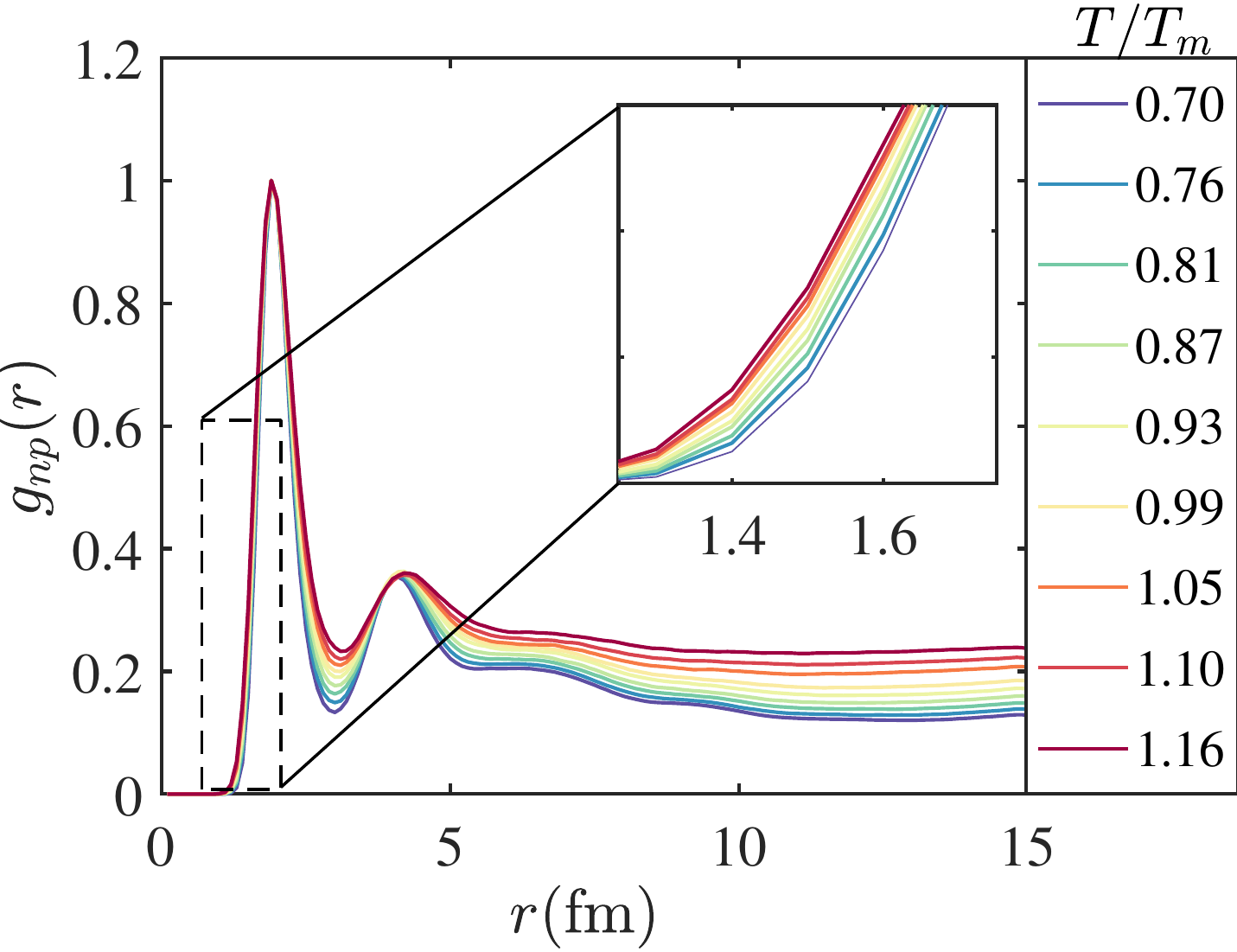}
\includegraphics[trim=100 240 100 235,clip,width=0.38\textwidth]{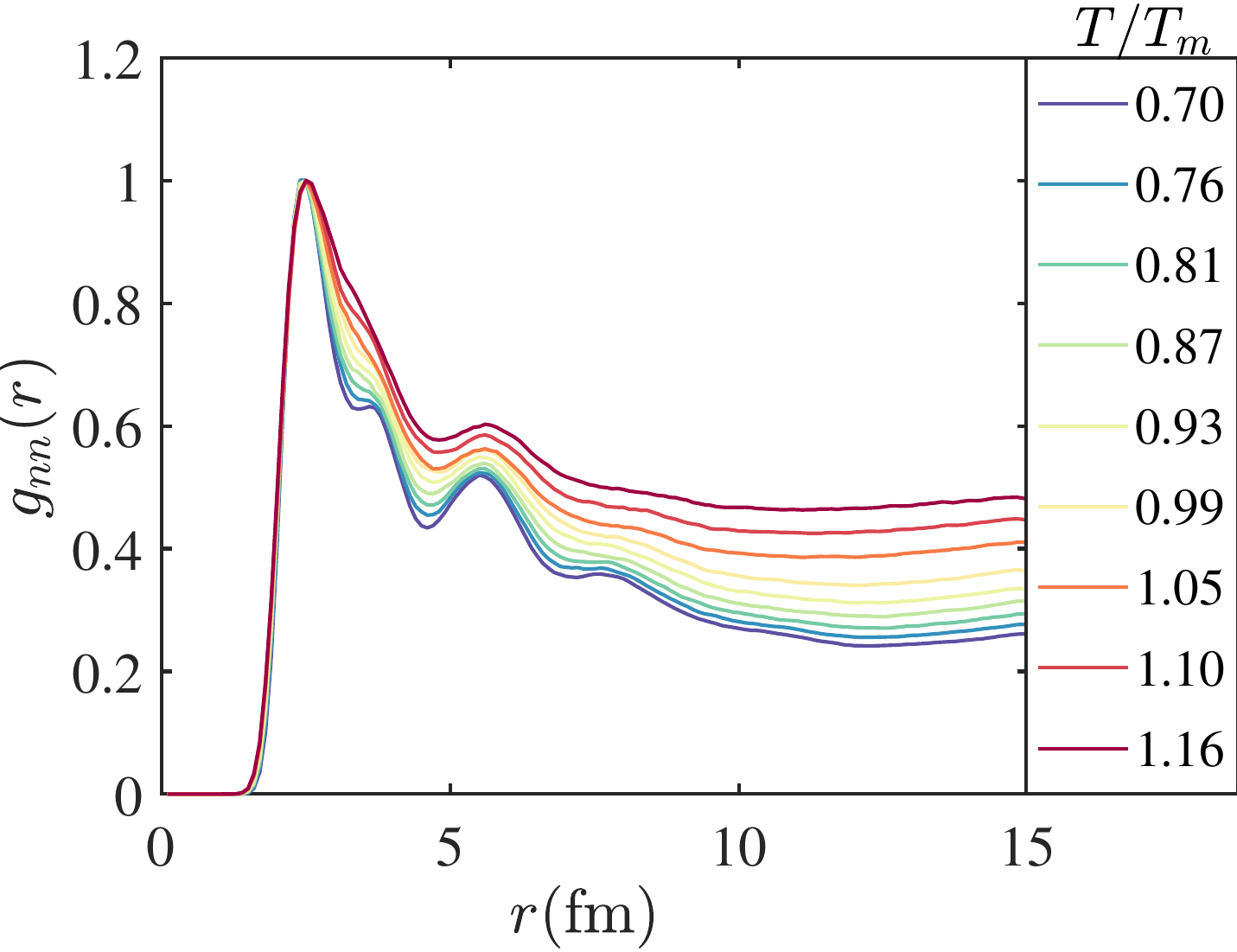}
\includegraphics[trim=100 240 100 235,clip,width=0.38\textwidth]{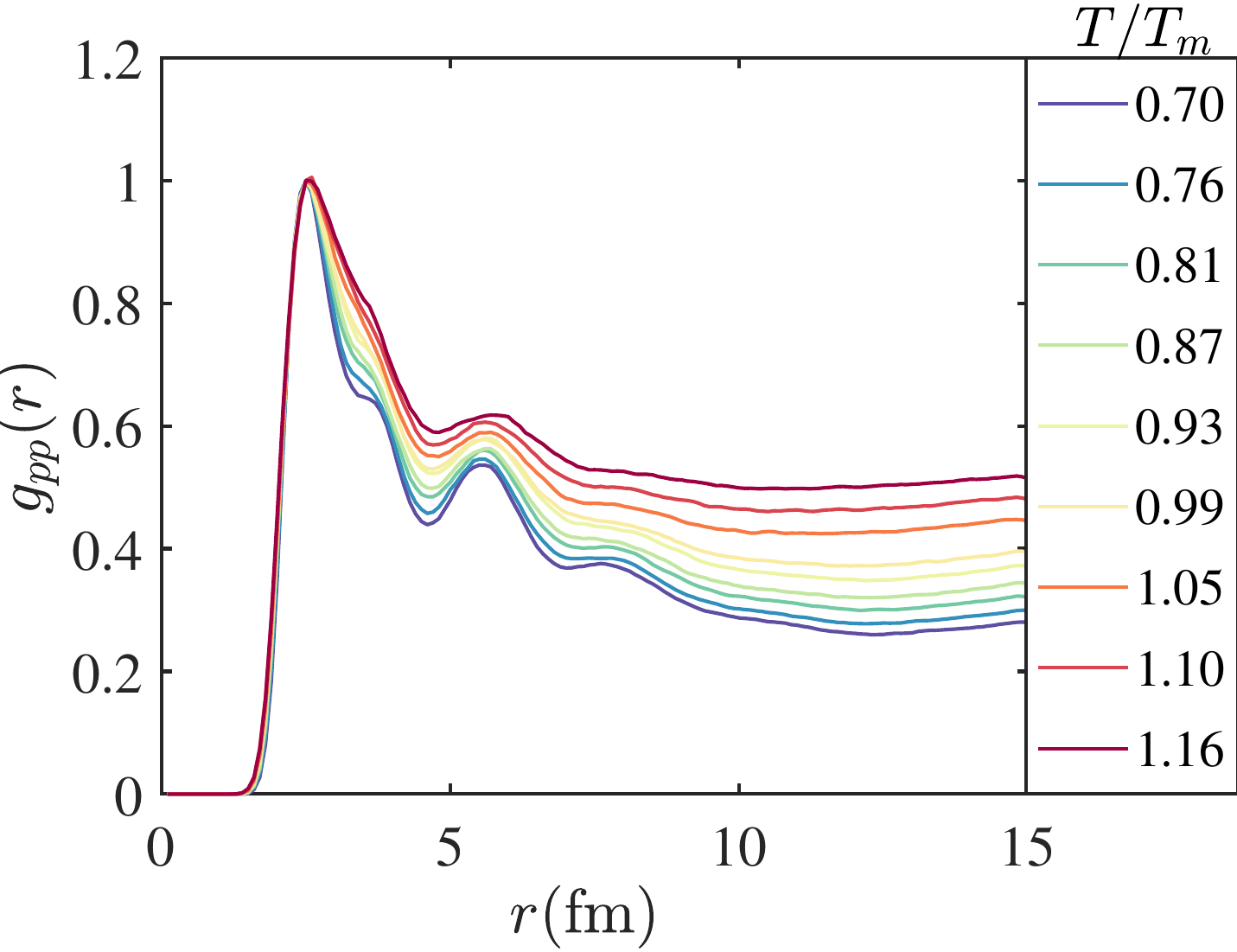}
\caption{\label{fig:gr} (Color online) Radial distribution functions $g(r)$ as function of temperature for (top) neutron-proton correlations, (center) neutron-neutron correlations, and (bottom) proton-proton correlations. We normalize to 1 at the position of the first peak  
which occurs at 1.9 fm for $g_{np}(r)$ and at 2.5 fm for both $g_{nn}(r)$ and $g_{pp}(r)$. The inset in $g_{np}(r)$ shows the approximate 0.1 fm broadening of the first peak, discussed in text. Below $T_m$ we use configurations from our `parallel' simulations. Note the similarity between $g_{nn}(r)$ and $g_{pp}(r)$ which may be due to the equal numbers of protons and neutrons in these simulations (compare to Fig. 2 in Ref. \cite{PhysRevC.88.065807})}	
\end{figure}


\subsubsection{Occupied Volume Fraction}

The occupied volume $V_{\mathrm{occ}}$ is the region bound by the gold and cream surfaces in our figures with proton density $n_{ch} > n_{th} = 0.03 \text{ fm}^{-3}$. The total volume  $V_{\mathrm{tot}}$ is just that of the cubic simulation volume. To zeroth order, approximately 41\% of the simulation volume contains condensed nuclear matter for all three topologies below the melting temperature.  At nucleon densities of 0.05 fm$^{-3}$, this suggests that uniform nuclear matter occurs around 0.12 fm$^{-3}$ at these proton fractions, which is consistent with other simulations used to produce phase diagrams of our pasta model \cite{PhysRevC.88.065807}. We observe that $V_{\mathrm{occ}} / V_{\mathrm{tot}}$ increases approximately linearly with temperature for $T/T_m \lesssim 0.8 $. We argue that this is due to greater average displacements of nucleons in the potential wells of nearest neighbors. Higher thermal velocities result in greater root mean square separations of nucleons producing slightly enlarged pasta structures, though the effect is small, of order $10^{-2}$. This can be seen clearly in the radial distribution functions $g(r)$ shown in Fig. \ref{fig:gr}. In the neutron-proton, neutron-neutron, and proton-proton pair correlations we see broadening of the first peak with temperature, with mean separations decreasing by about 0.1 fm when increasing $T/T_m$ from 0.70 to 0.87 (top inset). As the plate thickness and spacing are both nearly 10 fm, we see that this broadening explains the observed $\simeq10^{-2}$ enhancement in $V_{\mathrm{occ}} / V_{\mathrm{tot}}$. 

For $0.87 \lesssim T/T_m \lesssim 1.0 $ we observe a turnover in $V_{\mathrm{occ}} / V_{\mathrm{tot}}$. Naively this may seem to contradict our argument above, that broadening of the first peak in $g(r)$ with temperature should result in monotonically increasing $V_{\mathrm{occ}} / V_{\mathrm{tot}}$ with $T/T_m$. One possibility is that the mean square displacements may become sufficiently large that the mean nucleon density (near the surface) is below the threshold to count as being in the volume, \ie\ the surface becomes `puffy.' Additionally, some nucleons appear to be entering a sparse gas between plates, if this population is of order $10^{-2}N_{\mathrm{tot}}$, where $N_{\mathrm{tot}}$ is the number of nucleons in the simulation volume, then the reduction is explained. 

Above the melting temperature the occupied volume fraction shows a discontinuity consistent with a first order phase transition, and a steepening trend towards lower $V_{\mathrm{occ}} / V_{\mathrm{tot}}$ is observed with likely the same explanation (low density surfaces and losses of nucleons to the gas). This is again supported by $g(r)$; the loss of sharpness in the second peak and beyond suggests a more gas-like distribution of neighbors at $r>5$ fm, indicating that the characteristic pasta length-scale decreases with temperature above $T_m$, which can be seen in Fig. \ref{fig:heat} (j-l) as well as the SM.


\subsubsection{Surface Area}

Isosurfaces in charge density of $n_{ch} = n_{th} = 0.03$ fm$^{-3}$ are the gold surfaces in our figures. The total surface area (or equivalently, surface area density) increases with temperature for all three configurations and is discontinuous at $T/T_m=1.0$, consistent with a first order phase transition. This is easily explained by arguing that thermal fluctuations produce deviations in the surface such as filaments, holes, or buckling modes. Thermal fluctuations at greater temperatures produce greater average displacements of the surface, resulting in greater increases in surface area, and can clearly be seen in Fig \ref{fig:heat} (a-i). Observe that $A/V_\mathrm{tot}$ increases by about 20\% between $T/T_m = $ 0.70 to 0.99, comparable to the growth in temperature. 



\subsubsection{Mean Breadth}

We find that $B/A$ is monotonically increasing with temperature. At low temperature we observe different behavior for the three topologies. The mean breadth for the `nonparallel' and `parallel' configurations which contain only planar lasagna follow a power law ($B/A \propto  T^{9.7}$). The helicoidal ramps provide some nonzero surface curvature at low temperature, so that the `defects' configuration asymptotically approaches $B/A \approx 3 \times 10^{-3}$ at low temperatures. At temperatures approaching the melting temperature $B/A$ for the `defects' converges with what is seen in the `nonparallel' and `parallel' simulations, as thermal fluctuations come to dominate the surface curvature. It is worth noticing that the ‘nonparallel’ simulation at $0.86 T_m$ spontaneously forms small local helicoids, as discussed above. The calculated $B/A$ for this simulation is in closer agreement with that of the simulation with `defects' than the `parallel' simulation.  Asymptotic low $T/T_m$ behavior is similar to values obtained for simulations of same size and density but at lower proton fraction (Y=0.40) in previous works, $B/A \simeq 10^{-3}$ fm$^{-1}$, see Fig. 2 in ref. \cite{PhysRevC.93.065806} and Fig. 14 in ref. \cite{Schneider:18}. 


\subsubsection{Euler Characteristic}

First, observe the negative units of $\chi/A$ in Fig. \ref{fig:min} so that $\chi/A$ is actually monotonically decreasing. This indicates that the surfaces display saddle splay rather than convexity. As with the mean breadth there is a clear power law with temperature for the lasagna without helicoids ($-\chi/A \propto T^{12.5}$). The `defects' show the same behavior as in $B/A$; they asymptotically approach a nonzero value at low temperature due to the finite contribution to the curvature from the helicoids. The spontaneous formation of helicoids at $T/T_m = 0.87$ in the `nonparallel’ simulation again shifts $\chi/A$ for that run to become more in line with what is observed for the simulations with `defects.' As above, these results are same order of magnitude as for low $T$ runs in past work with $n = 0.05 \text{ fm}^{-3}$ and $Y_P=0.40$ which find $\chi/A \simeq 5\times10^{-5}$ fm$^{-2}$  \cite{PhysRevC.93.065806,Schneider:18}. 



\subsection{Static Structure Factors}\label{ssec:sq}

\begin{figure}[t!]
\centering
\includegraphics[trim=90 240 100 235,clip,width=0.38\textwidth]{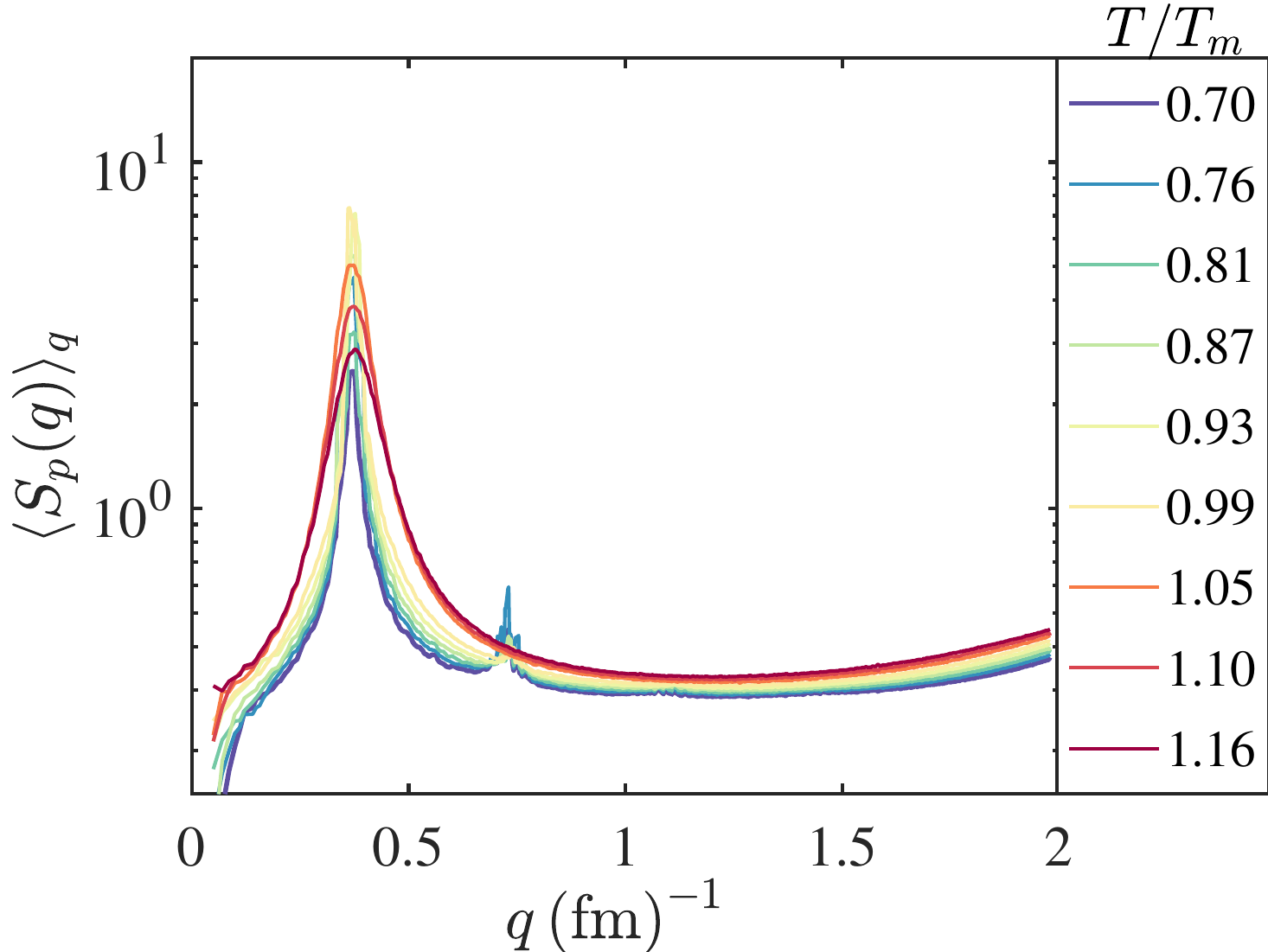}
\includegraphics[trim=90 240 100 235,clip,width=0.38\textwidth]{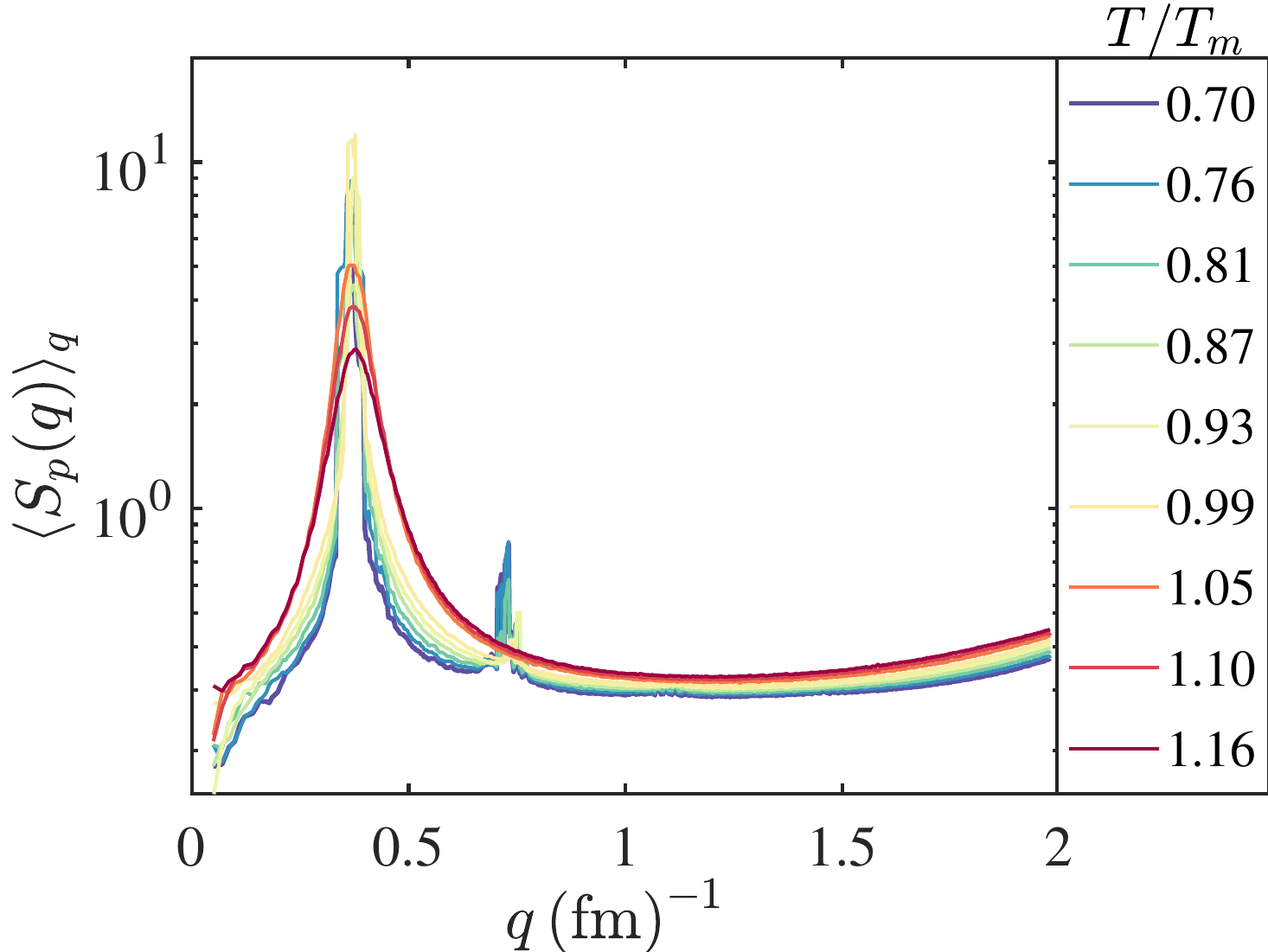}
\includegraphics[trim=90 240 100 235,clip,width=0.38\textwidth]{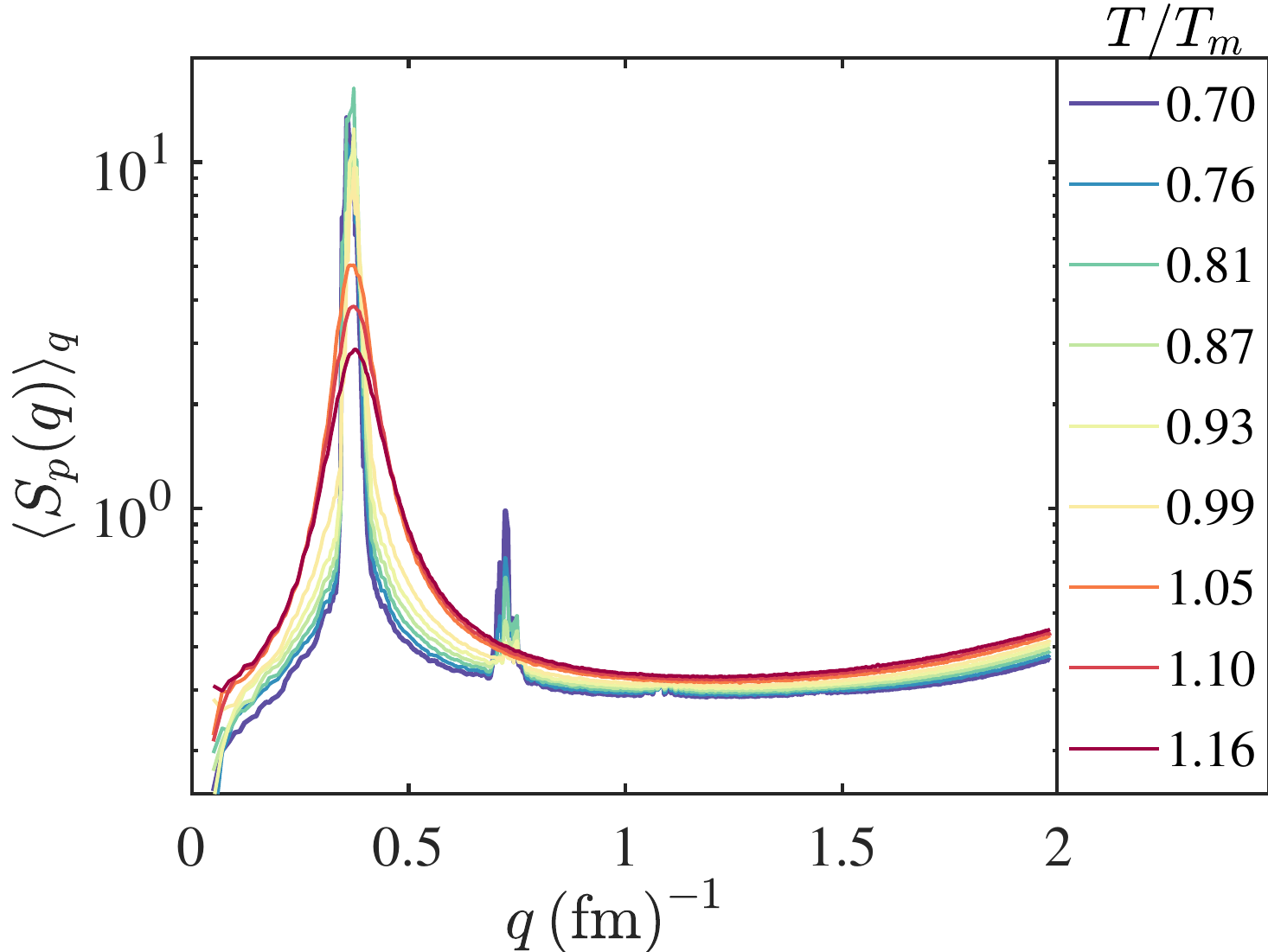}
\caption{\label{fig:sq} (Color online) Angle averaged proton structure factor $S_p(q) = \langle S_p ( \boldsymbol{q} ) \rangle_q$ for the range of temperatures studied, smoothed with Bragg peaks removed; (Top) configurations with defects, (middle) nonparallel plates, and (bottom) parallel plates. Lines $T/T_m > 1$ are the same in all three plots. Neutron structure factors $S_n(q)$ are nearly identical, following from similarities in $g_{nn}(r)$ and $g_{pp}(r)$ in Fig. \ref{fig:gr}. }	
\end{figure}

\begin{figure*}[t!]
\centering
\includegraphics[trim=0 0 10 10,clip,width=0.32\textwidth]{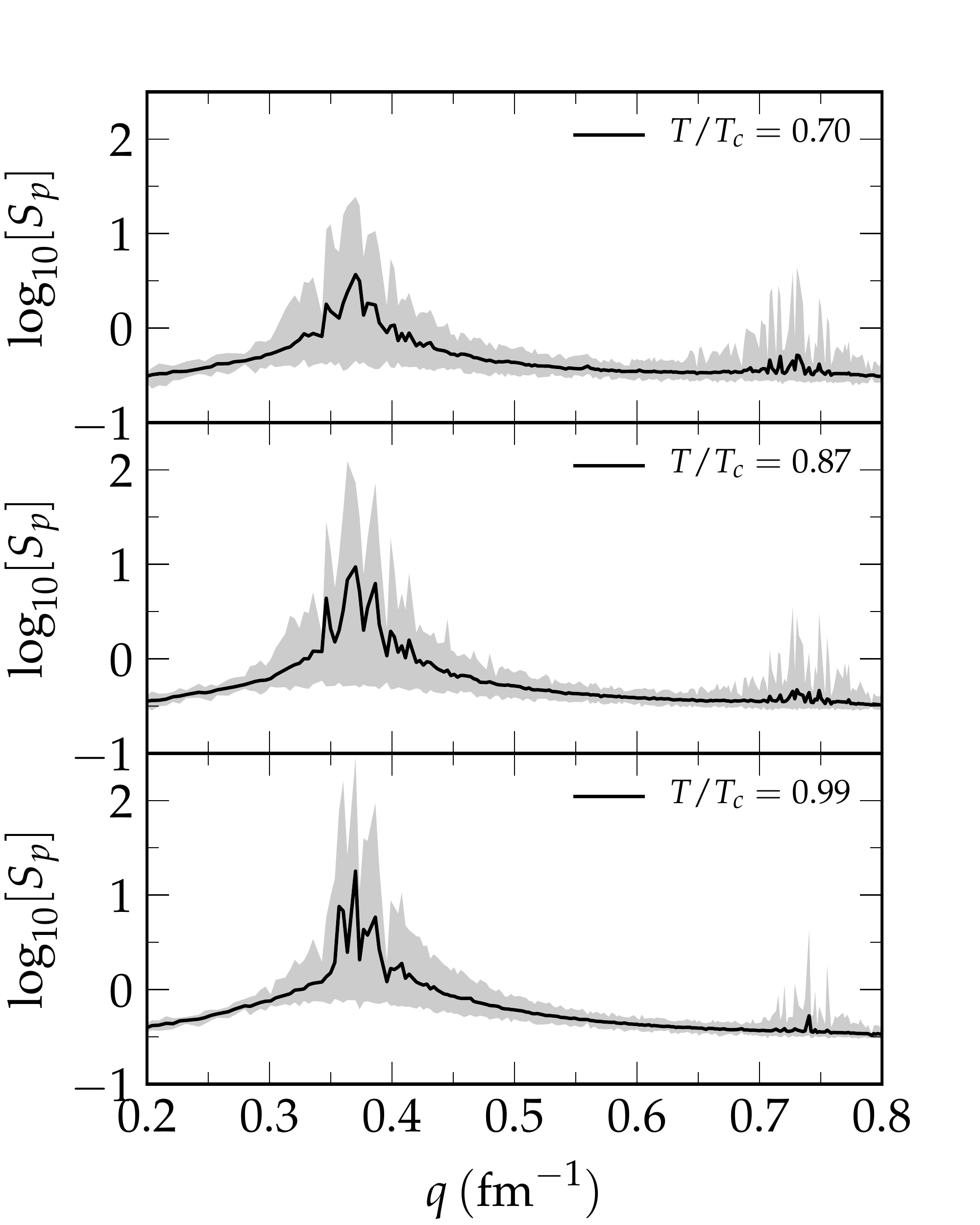}
\includegraphics[trim=0 0 10 10,clip,width=0.32\textwidth]{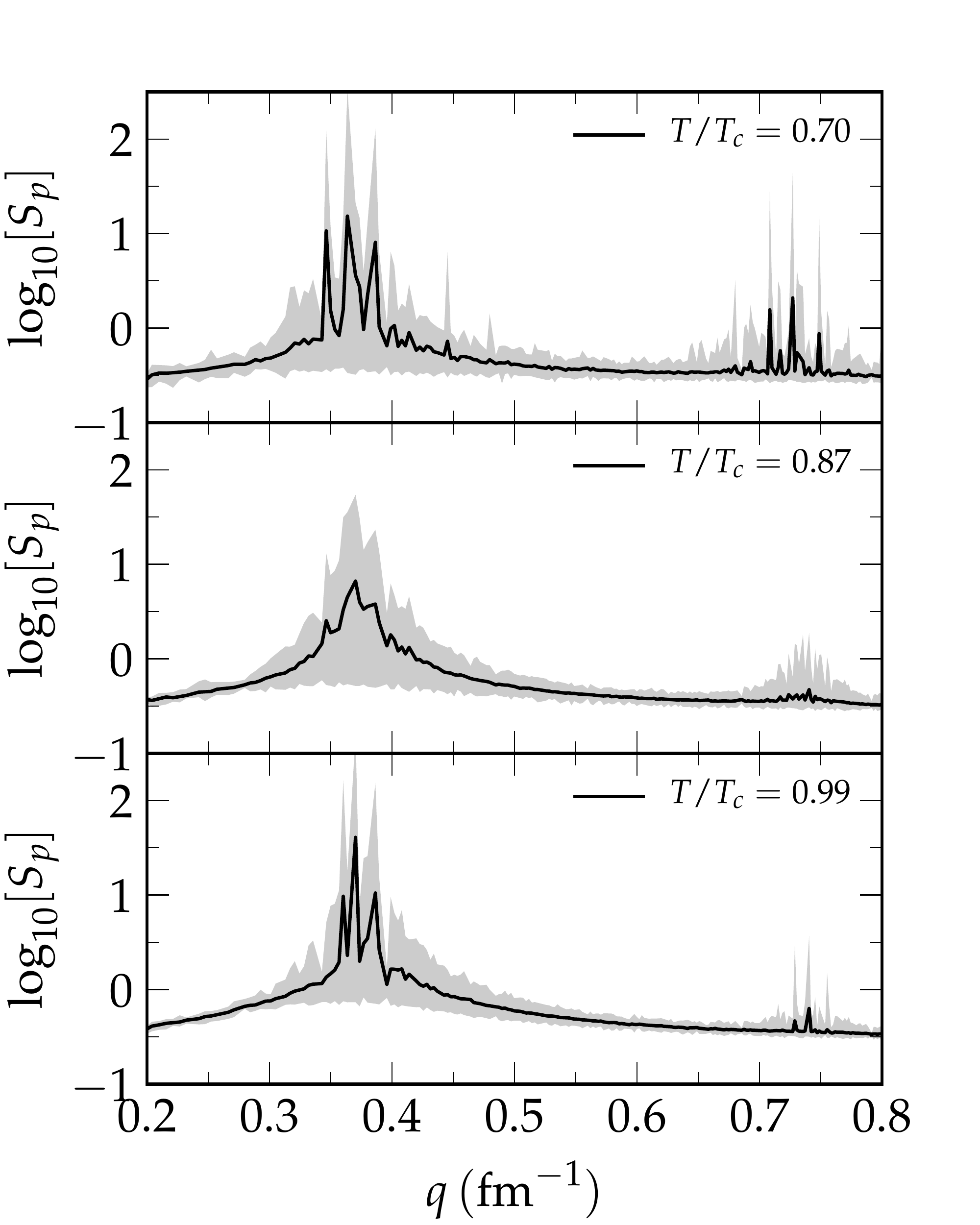}
\includegraphics[trim=0 0 10 10,clip,width=0.32\textwidth]{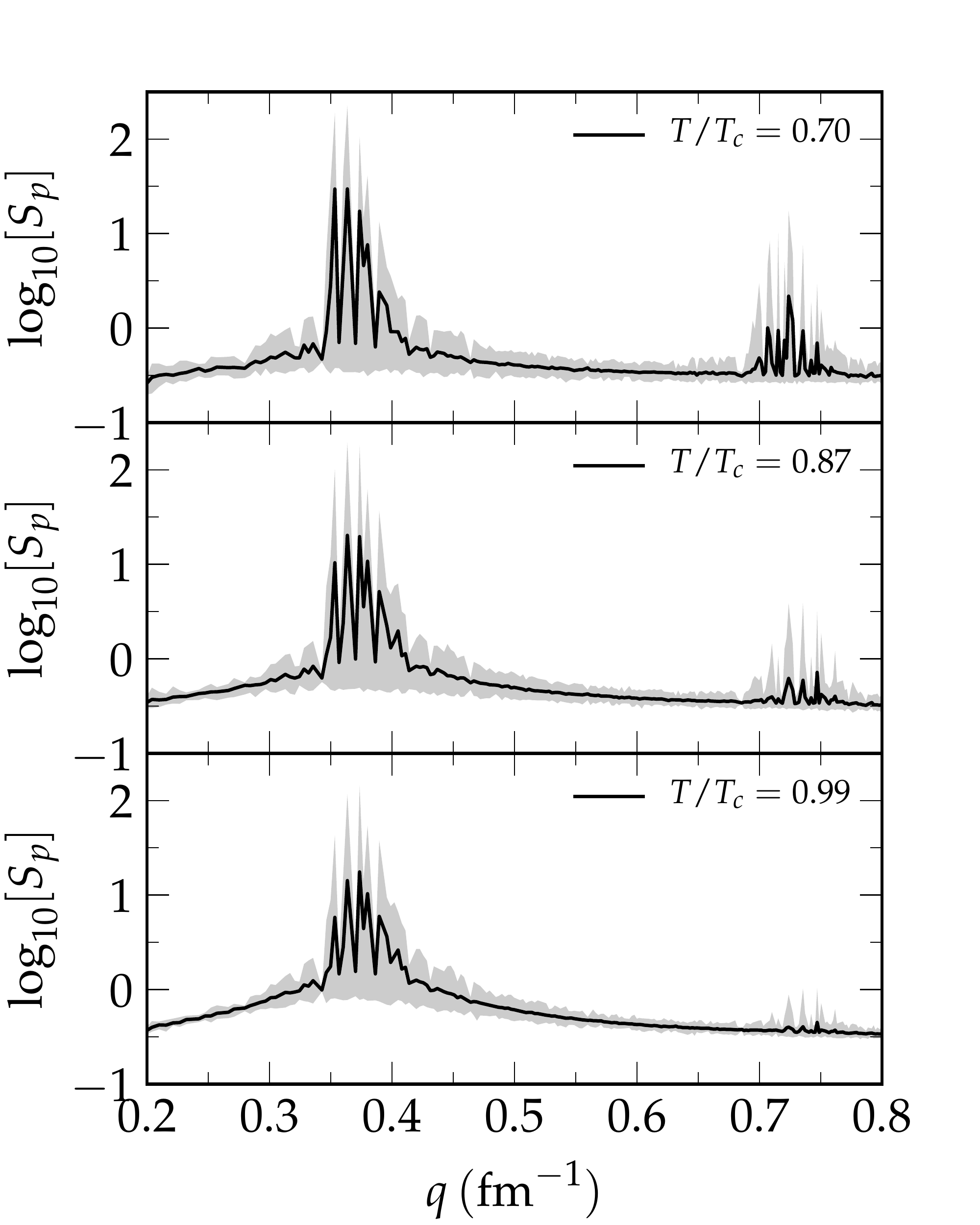}
\includegraphics[trim=0 0 10 10,clip,width=0.32\textwidth]{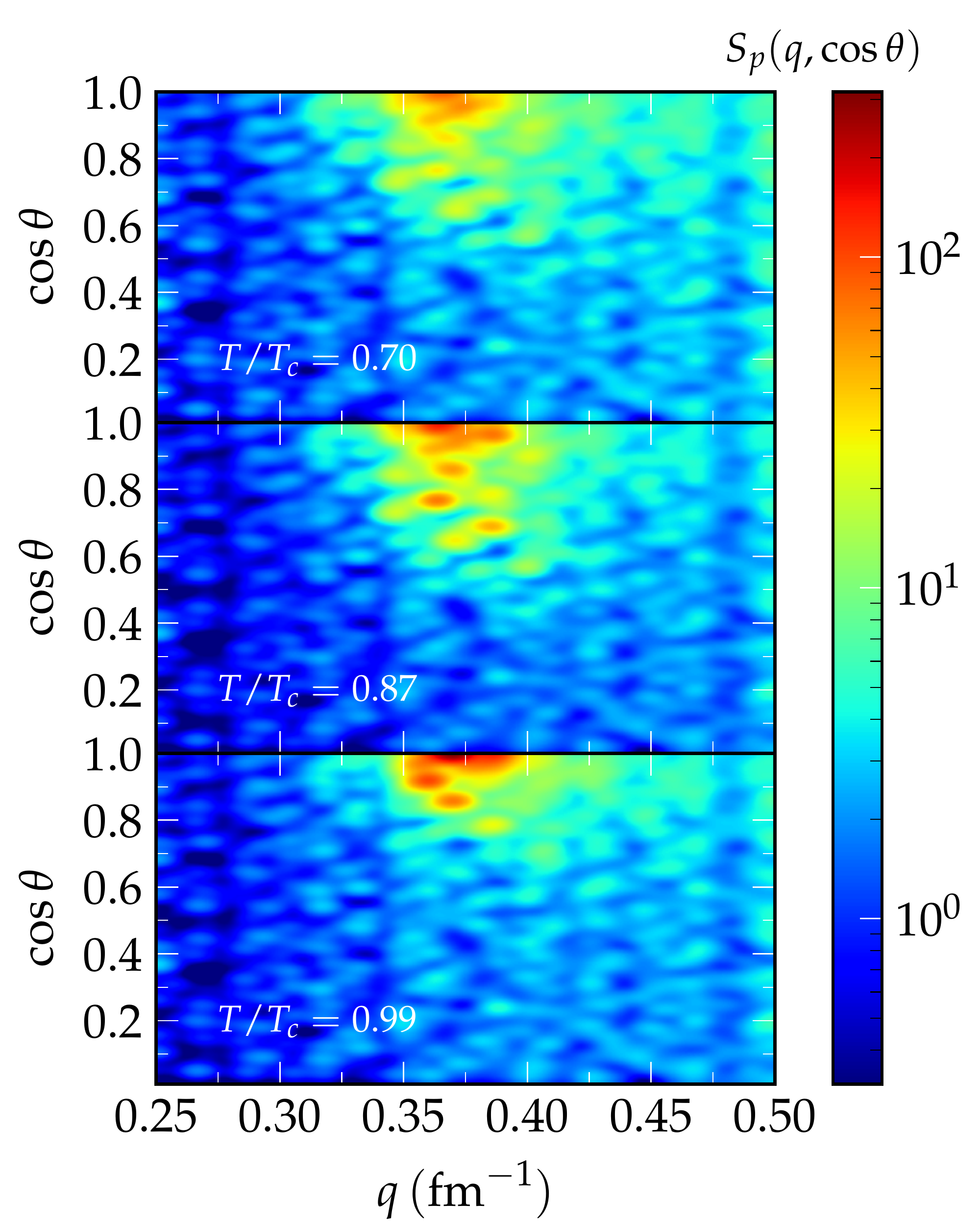}
\includegraphics[trim=0 0 10 10,clip,width=0.32\textwidth]{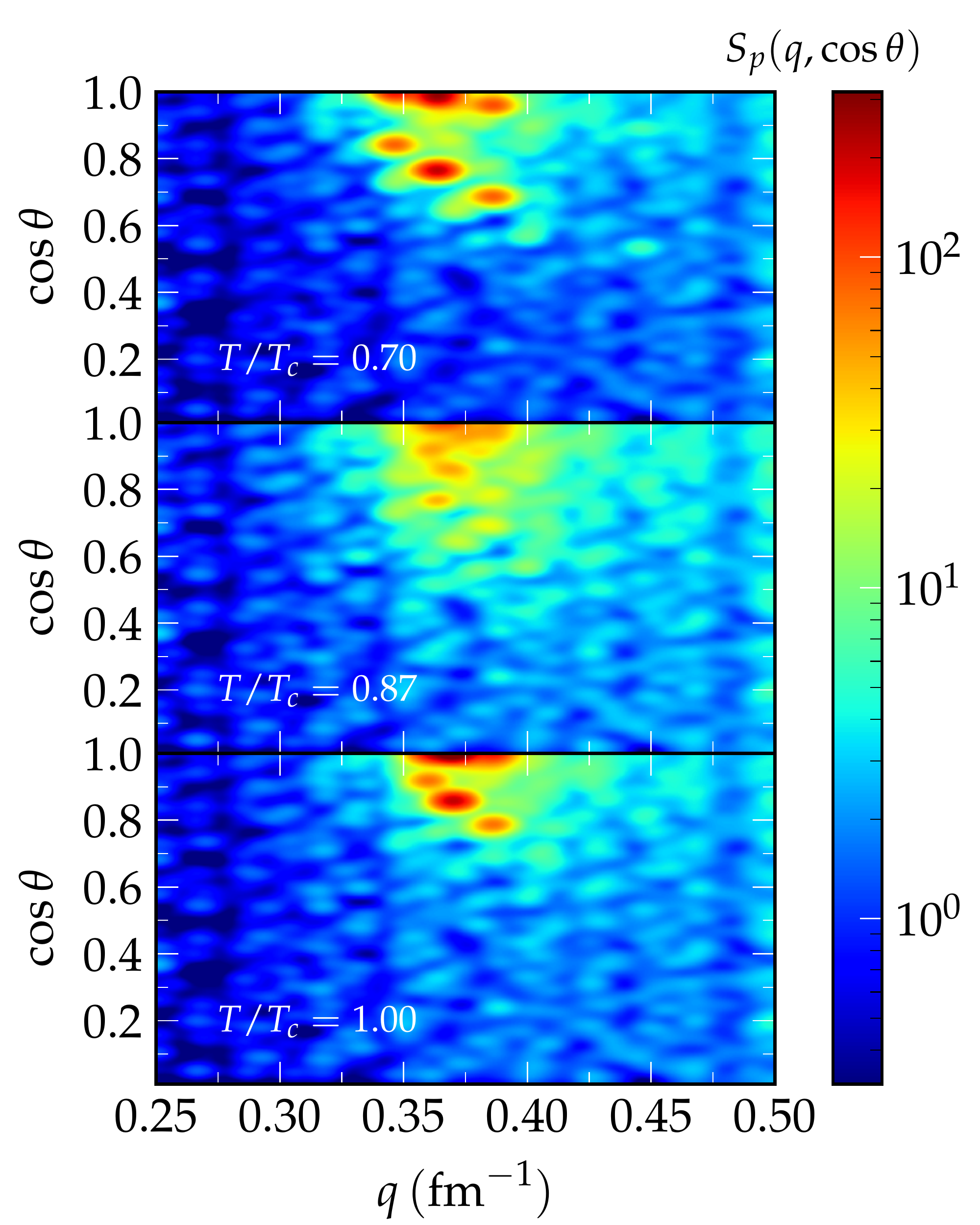}
\includegraphics[trim=0 0 10 10,clip,width=0.32\textwidth]{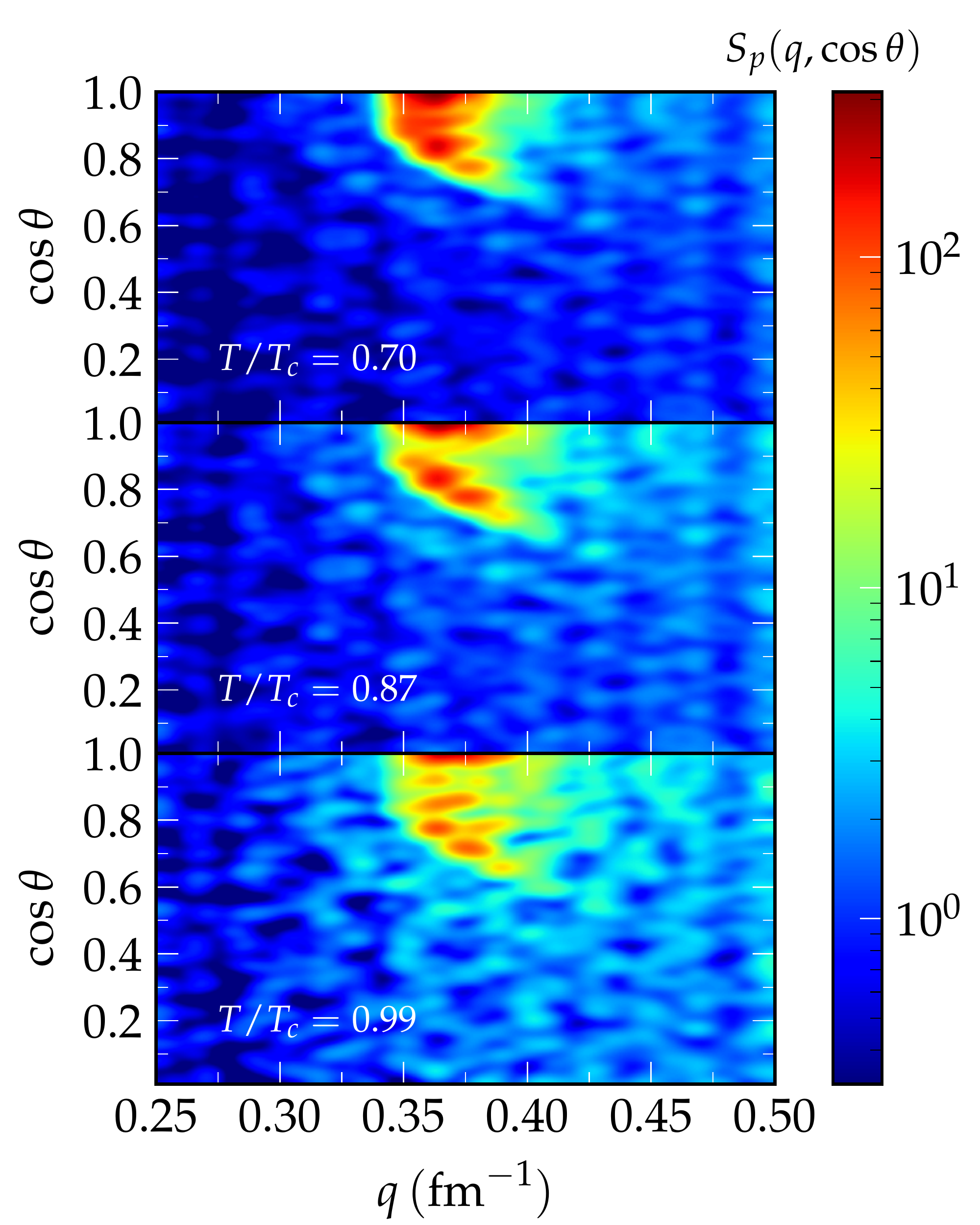}
\caption{\label{fig:sqT} (Color online) (Top) Angle averaged proton static structure factor $S_p(q) = \langle S_p ( \mathbf{q} ) \rangle_q$ (solid black) bounded by the maximum and minimum in $S_p(q)$ for each $q=| \mathbf{q} | $ for our simulations with (left) defects, (center) nonparallel plates, and (right) parallel plates. (Bottom) Heat map of proton structure factor $S_p(q)$ as a function of momentum transfer $q= |\mathbf{q}|$ and angle $\theta$. }	
\end{figure*}

\begin{figure}[t!]
\centering
\includegraphics[trim=0 0 10 10,clip,width=0.32\textwidth]{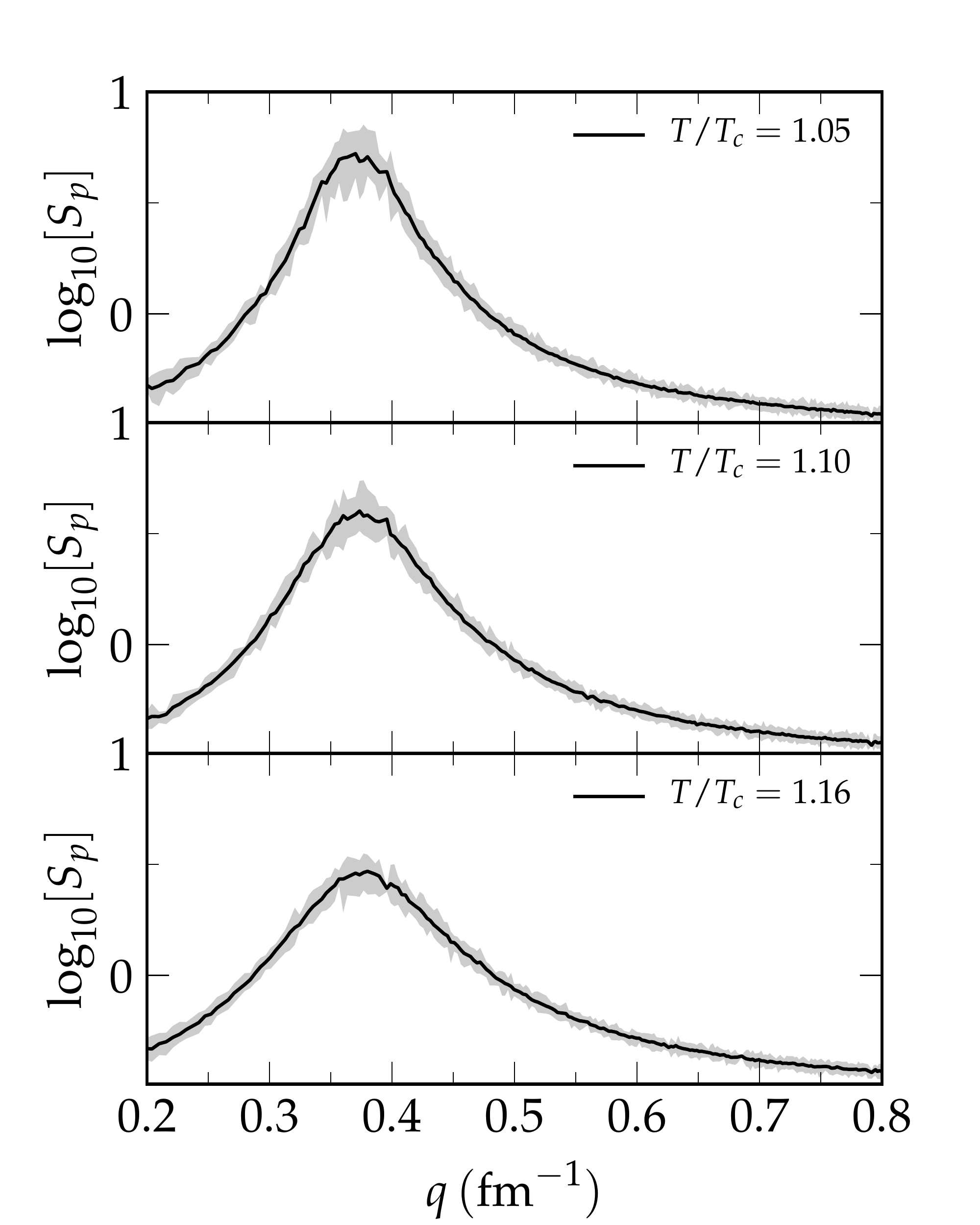}
\includegraphics[trim=0 0 10 10,clip,width=0.32\textwidth]{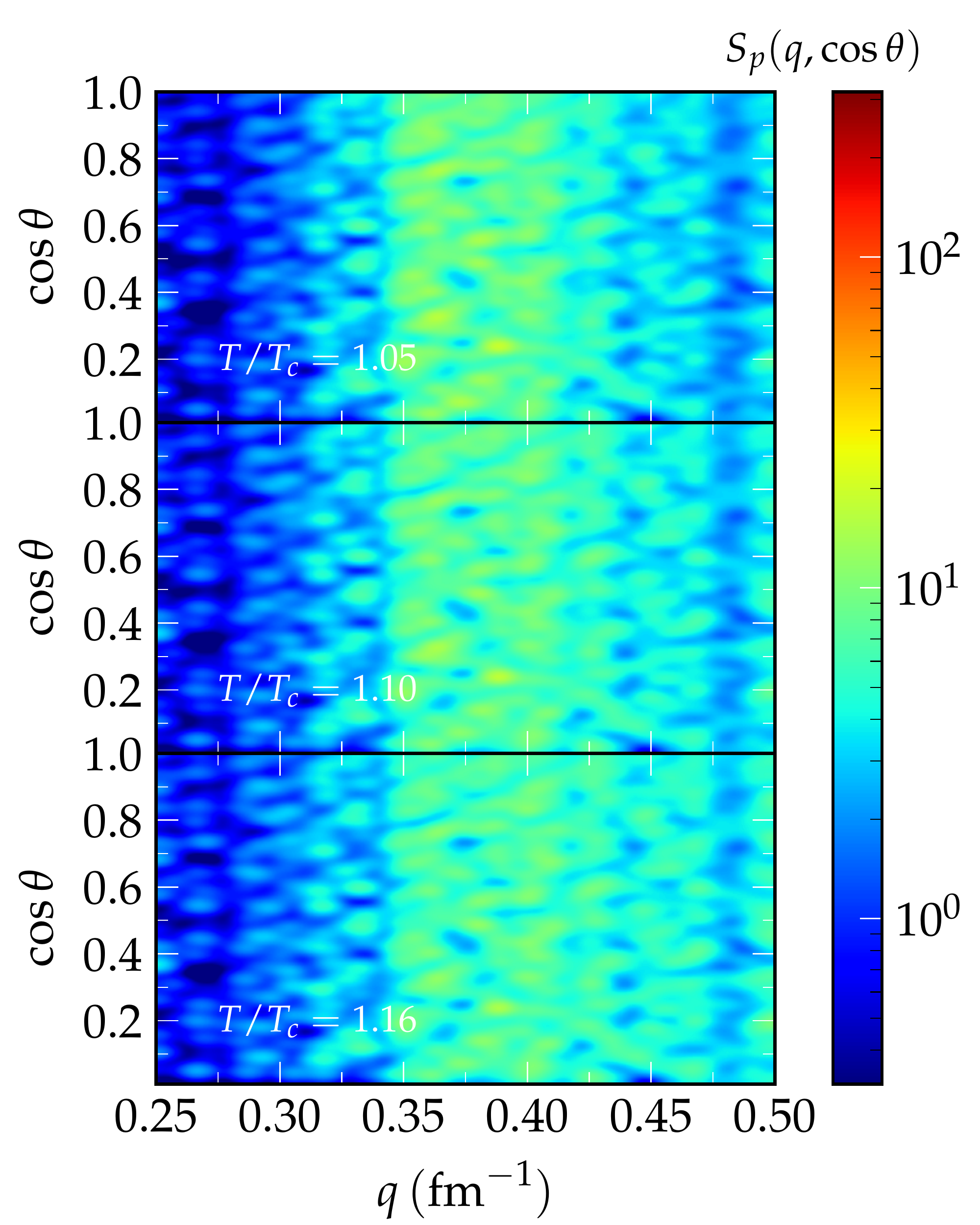}
\caption{\label{fig:sqT2} (Color online) (Top) Angle averaged proton static structure as in Fig. \ref{fig:sqT} for our disordered configurations above the melting temperature and (bottom) Heat map of proton structure factor $S_p(q)$ as a function of momentum transfer $q= |\mathbf{q}|$ and angle $\theta$.}
\end{figure}

We report on the static structure factor for nucleons in our simulations. As this is just the Fourier transform of the radial distribution function and our proton-proton and neutron-neutron radial distribution functions are nearly identical, we only report the proton static structure factors here. Our procedure for calculating these is described in detail in our past work (ref. \cite{PhysRevC.93.065806,Schneider:18}). Structure factors $S_p(\boldsymbol{q})$ are calculated from the time average (of $10^3$ MD configurations) of the nucleon density in momentum space:
\begin{equation}\label{eq:sq}
 S_p(\boldsymbol{q})=\langle\rho_p^*(\boldsymbol{q},t)\rho_p(\boldsymbol{q},t)\rangle_t
                    -\langle\rho_p^*(\boldsymbol{q},t)\rangle_t\langle\rho_p(\boldsymbol{q},t)\rangle_t
\end{equation}
with $\rho_p(\boldsymbol{q},t)=N_p^{-1/2}\sum_{j=1}^{N_p}e^{i\boldsymbol{q}\cdot\boldsymbol{r}_j(t)}$ 
the nucleon density in momentum space, $N_p$ the number of protons, and $\boldsymbol{r}_j(t)$ the position of the $j$-th proton at time $t$. The angled brackets $\langle{A}\rangle_t$ then denote the average of quantity $A$ over time interval $t$. 

Angle averaged proton static structure factors $S_p(q) = \langle S_p ( \boldsymbol{q}) \rangle$ are shown in Fig. \ref{fig:sq}. These $S_p(q)$ are smoothed to show the reduction and broadening of the first peak with temperature. As expected the static structure factor is largely independent of the exact configuration that we consider, but we do observe some small sensitivity in the magnitude of the first and second peaks which are sharpest in our parallel simulations and weakest in our simulations with defects. 

In Figs. \ref{fig:sqT} and \ref{fig:sqT2} we show detailed information about the static structure factors for three temperatures below and three temperatures above the melting temperature. 
In the top of Fig. \ref{fig:sqT} we show the angle averaged proton static structure factor, including the Bragg peaks composing the first maximum near $q \sim 0.37$ fm$^{-1}$ and second near $q \sim 0.75$ fm$^{-1}$. In the shaded regions we show the range between the  maximum and the minimum value of $S(q)$ for each $q$. These can be obtained from heatmaps similar to the ones shown in the bottom of Fig. 6, which show $S(q)$ for all $\theta$ and for $q$ near the first peak in $S(q)$.  Due to the finite box size only specific $(q,\theta)$ points can be calculated from which we interpolate to produce the heatmap, resulting in the apparent grainy texture. The interpolation and smoothing scheme is described in detail in our past work \cite{Schneider:18}.

Below the melting temperature we find that the structure of the peaks are largely independent of temperature, though we resolve a weak broadening of the peak with temperature as seen in Fig. \ref{fig:sq}. Most notably, the nonparallel plate configuration at $T=0.87 T_m$ shows the weakest Bragg peaks in the first peak. This is explained by the presence of small helicoidal defects with finite lifetimes which begin forming at this temperature. In contrast, the simulations above the melting temperature in Fig. \ref{fig:sqT2} show an order of magnitude reduction in the first peak relative to the configurations below the melting temperature. The first peak also decays in magnitude by approximately a factor of two over the temperatures studied. There is no apparent $\theta$ dependence observed above the melting temperature which is expected due to the relatively uniform randomness of the structure.

\subsection{Observables}\label{ssec:obs}

\begin{figure}[t!]
\centering
\includegraphics[trim=30 150 380 0,clip,width=0.49\textwidth]{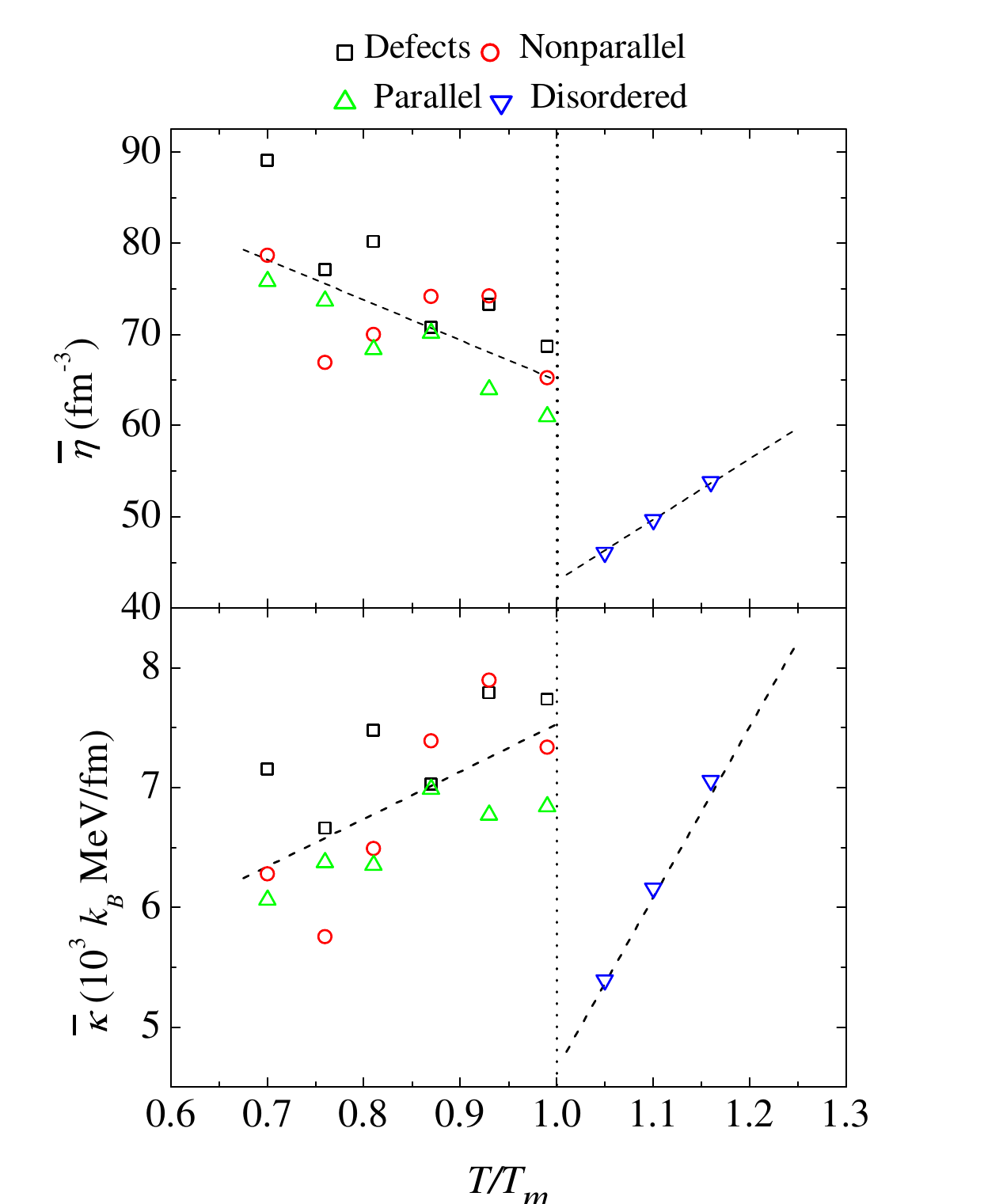}
\caption{\label{fig:obs} (Color online) Averaged shear viscosity (top) and averaged thermal conductivity (bottom).  
}	
\end{figure}

From the static structure factors we calculate the averaged shear viscosity $\bar{\eta}$ and thermal conductivity $\bar{\kappa}$, shown in Fig. \ref{fig:obs}. Following the methods of our previous work, we calculate 

\begin{equation}\label{eq:visc}
 \eta=\frac{\pi v^2_Fn_e}{20\alpha^2\Lambda^\eta_{\text ep}},
\end{equation}
\begin{equation}\label{eq:therm}
 \kappa=\frac{\pi v^2_F k_F k_B^2 T}{12\alpha^2\Lambda^\kappa_{\text ep}}
\end{equation}

\noindent using electron Fermi velocity and momentum $v_F$ and $k_F$, electron density $n_e$, fine structure and Boltzmann constants $\alpha$ and $k_B$, with $T$ the temperature of the system \cite{PhysRevC.93.065806}. We approximate the Coulomb logarithms $\Lambda^\eta_{ep}$ and $\Lambda^\kappa_{ep}$ 
via 

\begin{equation}\label{eq:visc2}
 \Lambda^\eta_{\text ep}=\int_0^{2k_F}\frac{dq}{q\epsilon^2(q)}\left(1-\frac{q^2}{4k_F^2}\right)\left(1-\frac{v_F^2q^2}{4k_F^2}\right)S_p(q)
\end{equation} 
\begin{equation}\label{eq:therm2}
 \Lambda^\kappa_{\text ep}=\int_0^{2k_F}\frac{dq}{q\epsilon^2(q)}\left(1-\frac{v_F^2q^2}{4k_F^2}\right)S_p(q).
\end{equation}

\noindent where $\epsilon(q)$ is the Thomas-Fermi approximation to the dielectric function, taken to be $\epsilon(q)=1+k_{TF}^2/q^2$ which uses the inverse screening length $k_{TF}  \equiv \lambda^{-1}=2k_F\sqrt{\alpha/\pi}$. We calculate $k_F=(3\pi^2n_e)^{1/3}$ from the electron (proton) density $n_e$ by assuming charge neutrality. We use $k_{TF}^{-1} = 11.5$ fm \footnote{In our simulations we use $\lambda = 10$ fm for the proton-proton Coulomb screening. However, using $k_{TF}^{-1} = 11.5$ or $k_{TF}^{-1} = 10$ results in only a 2\% variation in $\Lambda_{ep}$.}\cite{PhysRevC.93.065806,PhysRevC.90.055805}. Lastly, the angle averaged $\bar{\eta}$ is found by 
\begin{equation}\label{eq:avg}
 \bar{\eta}=\frac{\int \eta(\theta)\sin\theta d\theta}{\int \sin\theta d\theta}.
\end{equation}
\noindent and similarly for $\bar{\kappa}$.


Our results for the shear viscosity $\bar{\eta}$ and the thermal conductivity $\bar{\kappa}$ are of the same order as in our past work and we resolve rough trends with temperature \cite{PhysRevC.88.065807}. These results are about one order of magnitude larger than the ones obtained by Nandi and Schramm \cite{Nandi2018} considering the same proton fraction, $Y_p=0.5$, temperatures in the range from 0 to 5 MeV in 1 MeV increments, and similar densities, $\rho/\rho_0=0.3$ and $0.4$\footnote{Although $\rho/\rho_0=0.3$ better matches the density we simulate in this work, $\rho/\rho_0=0.4$ is where the QMD model often finds the lasagna phase \cite{PhysRevC.66.012801, PhysRevC.68.035806}. Therefore, we look at both densities when making parallels between our results and those of Ref. \cite{Nandi2018}.} with the nuclear saturation density $\rho_0=0.165\,{\rm fm}^{-3}$ \cite{PhysRevC.57.655}. We atribute this order of magnitude difference to the smaller simulation sizes of Nandi and Schramm as their runs contained 12288 nucleons. Smaller simulation volumes can increase correlations between nucleons in the pasta due to the periodic boundary conditions, leading to higher peaks in $S_p(q)$ and, thus, larger Coulomb logarithms which appear in the denominator of Eqs. \eqref{eq:visc} and \eqref{eq:therm}. 

We observe in our simulations that the ‘parallel’ configurations show a fairly linear trend in both $\bar{\eta}$ and $\bar{\kappa}$, which we argue most reasonably captures the evolution of the observables with temperature. The large fluctuations of the ‘defects’ and ‘nonparallel’ simulations are due to spontaneous formation and dissolution of defects which biases our averaging when calculating $S(q)$, as shown in Sec. \ref{ssec:sq}. Coarsely, we can at least see the approximate trend of $\bar{\eta}$ decreasing with temperature below $T/T_m=1$ and $\bar{\kappa}$ increasing with temperature below $T/T_m=1$. A discontinuity in both $\bar{\eta}$ and $\bar{\kappa}$ are consistent with the first order phase transition at $T/T_m=1$, where both $\bar{\eta}$ and $\bar{\kappa}$ drop by about 30-40\% before increasing again. Given how $S(q)$ behaves for $T/T_m>1$ (Fig. \ref{fig:sqT2}), where there is not much noise in the angle averaged static structure factor, the time averaged $S(q)$ is more precise above the melting temperature than below. 

Finally, we note here the QMD formalism from Maruyama \textit{et al.} \cite{PhysRevC.57.655} and used by Nandi and Schramm \cite{PhysRevC.94.025806, PhysRevC.95.065801, Nandi2018} allows pasta structures to exist at higher temperatures than in our semi-classecal MD simulations. In runs that explore a similar parameter space to ours in Ref. \cite{Nandi2018}, we infer from the decrease in $S_p(q)$ that pasta structures melt between 3 and 4 MeV for $Y_p=0.5$, see also Ref. \cite{PhysRevC.95.065801}. 
However, due to the large 1 MeV increments in temperature, it is not clear in Ref. \cite{Nandi2018} what type of phase transition takes place as the pasta melts, although it is argued in Ref. \cite{PhysRevC.95.065801} for a $Y_p=0.30$ system that the transition observed is also of first order, nor if the overall topology of the pasta is similar across the range of temperatures explored. 
Still, Nandi and Schramm determine that the thermal conductivity increases fast with temperature below 5 MeV while the shear viscosity shows no clear temperature dependence for $Y_p=0.50$ and $\rho/rho_0=0.3-0.4$.
We speculate that the discrepancies in observed trends for the thermal conductivity between Ref. \cite{Nandi2018} and our results are due to finite size effects and differences in pasta topology.

Under more realistic circumstances (like what has been seen larger MD simulations, see refs. \cite{CaplanPRL,Schneider:18}) it is reasonable to expect features like transient defects and domains below the melting temperature, so the simulations containing defects with finite lifetimes do not necessarily give us unphysical results. We show a linear fit, motivated by the nearly smooth trend with $T/T_m$ seen in our `parallel' configurations, which effectively allows us to average over the kind of structures seen in all of our simulations below the melting temperature. Far below the melting temperature we expect $\bar{\eta}$ and $\bar{\kappa}$ to have different asymptotic behavior as they converge to values characteristic of cold catalyzed nuclear pasta.






\section{Discussion}\label{sec:dis}

We have resolved the behavior of nuclear pasta at a range of finite temperatures. In a cooling neutron star one might expect nuclear pasta to evolve through these phases, which may determine the ground state structure of pasta once annealed.
Astrophysical cooling mechanisms operate on much longer timescales than the characteristic nuclear timescales in pasta. Therefore, in an annealing neutron star crust, the pasta might be expected to be in a quasi-equilibrium state at any given temperature above some quenching temperature. The temperature that the topological thermal fluctuations are quenched out may be an effective freezing temperature for the nuclear pasta layer in neutron star crusts which determines the domain size and thus transport properties.

Consider the geometric evolution of a volume of subsaturation density matter in a cooling neutron star. Below the critical temperature nuclear pasta can form but it has many short lived topological defects such as holes and filaments. These filaments may provide a mechanism for annealing the crust by exchanging nucleons between plates. At lower temperatures, we observe that these filaments and holes can be more organized in the form of large helicoidal defects. We observe both the spontaneous dissolution and formation of helicoidal defects in simulations at the same temperature which suggests there is a critical temperature for their formation. These helicoids appear to be metastable at high temperature, and may be frozen in as the pasta anneals; energy differences between similar shapes may be small and timescales for tunneling may be large given the large number of nucleons involved. Once frozen in they may interact weakly via a long range attractive force causing them to cluster into dipoles or quadrupoles of alternating handedness (see refs. \cite{PhysRevC.94.055801,PhysRevLett.113.188101}). 

In contrast, short lived topological fluctuations at high temperatures may be a mechanism to anneal pasta and relieve stress via creep. 
Even if tunneling barriers between similar pasta structures are large, as in a glass, stress may be relieved by slowly exchanging nucleons between plates and changing the topology. We speculate that there may be some temperature threshold where filaments and holes may form on timescales comparable to astrophysical cooling, potentially relaxing the crust. Their presence may allow for the probing of many different pathways through the energy landscape and allow the pasta structures to reach lower energy, lower stress states. Therefore, topological thermal fluctuations may provide a mechanism to relieve stress. 

This work also observes evolution of the plate splay/buckling with increasing temperature, which may similarly affect the elastic properties. In Caplan \etal\ \cite{CaplanPRL}, we argue that that the `defects' can produce large shear moduli in the pasta, while parallel plates of the `lasagna' phase have zero in-plane shear modulus, as plates may slide freely parallel to each other. Dissolution of the helicoids at high temperature may effectively weaken the pasta, but high temperature non-topological thermal fluctuations may also stiffen the pasta. Surface roughness of the plates may provide some resistance to sliding \cite{pethick2019dense,pethick2020elastic}. As we observe that there may generally be some spontaneous curvature of the pasta surfaces, for example hyperbolic splay, one might expect corrugations to resist to shear stresses even at high temperature. How the magnitude of such shear modulus compares to the topological shear modulus studied in ref. \cite{CaplanPRL} remains to be seen, but taken together this motivates future work studying the thermoelastic properties of pasta. 

Bridging equations of state from the crust to the core will likely require corrections at subsaturation density for pasta. While some of the exact results in this work are model dependent (such as the occupied volume fraction and exact melting temperature) and are perhaps less useful for astrophysics, others may be general features of a liquid drop model for pasta. The nuclear pasta model in this work has been fit to reproduce known properties of nuclear matter near saturation, and should be expected to reproduce at least the bulk behavior of the pasta structure in the classical limit of many thousands of nucleons. For example, the surface area density found in this work may be useful for developing surface energy corrections to equations of state at pasta densities which bridge nuclear equations of state to the ion crust above it. Similarly, the observed surface roughness could motivate the inclusion of next-order surface energy terms, such as a curvature term, similar to curvature energy corrections used for models of fission and permanent nuclear deformations \cite{PhysRevC.73.014309,PhysRevC.83.065811,PhysRev.89.1102}. 

The observables we report show interesting evolution with temperature near the melting temperature. Given the large proton fractions used and the small sizes of the simulations reported in this work the exact values of  $\bar{\eta}$ and $\bar{\kappa}$ we report have considerable uncertainty. However, the apparent trends may be useful in astrophysical simulations where nuclear matter reaches high temperature. Discontinuities at the melting temperature could have interesting astrophysical implications, especially since our results suggest the viscosity reaches its minimum at the melting temperature. Detailed calculations of the observables as a function of temperature may not be easily accessible to MD without large simulations and long simulation times. Thus, this motivates future work which goes beyond MD to model pasta in a more computationally efficient manner, like a scalar field models and others common in the diblock copolymer literature \cite{rumyantsev2020microphase}.


\textit{Acknowledgements} The authors thank C. J. Horowitz and Z. Lin for conversation and Indiana University for hospitality. This work was enabled in part by the National Science Foundation under Grant No. PHY-1430152 (JINA Center for the Evolution of the Elements). This research was supported in part by Lilly Endowment, Inc., through its support for the Indiana University Pervasive Technology Institute, and in part by the Indiana METACyt Initiative. The Indiana METACyt Initiative at IU was also supported in part by Lilly Endowment, Inc. This material is based upon work supported by the National Science Foundation under Grant No. CNS-0521433. This work was supported in part by Shared University Research grants from IBM, Inc., to Indiana University.


\bibliography{bibliography}

\providecommand{\noopsort}[1]{}\providecommand{\singleletter}[1]{#1}%
\begin{thebibliography}{41}%
\makeatletter
\providecommand \@ifxundefined [1]{%
 \@ifx{#1\undefined}
}%
\providecommand \@ifnum [1]{%
 \ifnum #1\expandafter \@firstoftwo
 \else \expandafter \@secondoftwo
 \fi
}%
\providecommand \@ifx [1]{%
 \ifx #1\expandafter \@firstoftwo
 \else \expandafter \@secondoftwo
 \fi
}%
\providecommand \natexlab [1]{#1}%
\providecommand \enquote  [1]{``#1''}%
\providecommand \bibnamefont  [1]{#1}%
\providecommand \bibfnamefont [1]{#1}%
\providecommand \citenamefont [1]{#1}%
\providecommand \href@noop [0]{\@secondoftwo}%
\providecommand \href [0]{\begingroup \@sanitize@url \@href}%
\providecommand \@href[1]{\@@startlink{#1}\@@href}%
\providecommand \@@href[1]{\endgroup#1\@@endlink}%
\providecommand \@sanitize@url [0]{\catcode `\\12\catcode `\$12\catcode
  `\&12\catcode `\#12\catcode `\^12\catcode `\_12\catcode `\%12\relax}%
\providecommand \@@startlink[1]{}%
\providecommand \@@endlink[0]{}%
\providecommand \url  [0]{\begingroup\@sanitize@url \@url }%
\providecommand \@url [1]{\endgroup\@href {#1}{\urlprefix }}%
\providecommand \urlprefix  [0]{URL }%
\providecommand \Eprint [0]{\href }%
\providecommand \doibase [0]{http://dx.doi.org/}%
\providecommand \selectlanguage [0]{\@gobble}%
\providecommand \bibinfo  [0]{\@secondoftwo}%
\providecommand \bibfield  [0]{\@secondoftwo}%
\providecommand \translation [1]{[#1]}%
\providecommand \BibitemOpen [0]{}%
\providecommand \bibitemStop [0]{}%
\providecommand \bibitemNoStop [0]{.\EOS\space}%
\providecommand \EOS [0]{\spacefactor3000\relax}%
\providecommand \BibitemShut  [1]{\csname bibitem#1\endcsname}%
\let\auto@bib@innerbib\@empty
\bibitem [{\citenamefont {Schuetrumpf}\ \emph {et~al.}(2013)\citenamefont
  {Schuetrumpf}, \citenamefont {Klatt}, \citenamefont {Iida}, \citenamefont
  {Maruhn}, \citenamefont {Mecke},\ and\ \citenamefont
  {Reinhard}}]{schuetrumpf2013time}%
  \BibitemOpen
  \bibfield  {author} {\bibinfo {author} {\bibfnamefont {B.}~\bibnamefont
  {Schuetrumpf}}, \bibinfo {author} {\bibfnamefont {M.~A.}\ \bibnamefont
  {Klatt}}, \bibinfo {author} {\bibfnamefont {K.}~\bibnamefont {Iida}},
  \bibinfo {author} {\bibfnamefont {J.}~\bibnamefont {Maruhn}}, \bibinfo
  {author} {\bibfnamefont {K.}~\bibnamefont {Mecke}}, \ and\ \bibinfo {author}
  {\bibfnamefont {P.-G.}\ \bibnamefont {Reinhard}},\ }\href@noop {} {\bibfield
  {journal} {\bibinfo  {journal} {Physical Review C}\ }\textbf {\bibinfo
  {volume} {87}},\ \bibinfo {pages} {055805} (\bibinfo {year}
  {2013})}\BibitemShut {NoStop}%
\bibitem [{\citenamefont {Caplan}\ and\ \citenamefont
  {Horowitz}(2017{\natexlab{a}})}]{astromaterials}%
  \BibitemOpen
  \bibfield  {author} {\bibinfo {author} {\bibfnamefont {M.~E.}\ \bibnamefont
  {Caplan}}\ and\ \bibinfo {author} {\bibfnamefont {C.~J.}\ \bibnamefont
  {Horowitz}},\ }\href {\doibase 10.1103/RevModPhys.89.041002} {\bibfield
  {journal} {\bibinfo  {journal} {Rev. Mod. Phys.}\ }\textbf {\bibinfo {volume}
  {89}},\ \bibinfo {pages} {041002} (\bibinfo {year}
  {2017}{\natexlab{a}})}\BibitemShut {NoStop}%
\bibitem [{\citenamefont {Schneider}\ \emph {et~al.}(2019)\citenamefont
  {Schneider}, \citenamefont {Constantinou}, \citenamefont {Muccioli},\ and\
  \citenamefont {Prakash}}]{PhysRevC.100.025803}%
  \BibitemOpen
  \bibfield  {author} {\bibinfo {author} {\bibfnamefont {A.~S.}\ \bibnamefont
  {Schneider}}, \bibinfo {author} {\bibfnamefont {C.}~\bibnamefont
  {Constantinou}}, \bibinfo {author} {\bibfnamefont {B.}~\bibnamefont
  {Muccioli}}, \ and\ \bibinfo {author} {\bibfnamefont {M.}~\bibnamefont
  {Prakash}},\ }\href {\doibase 10.1103/PhysRevC.100.025803} {\bibfield
  {journal} {\bibinfo  {journal} {Phys. Rev. C}\ }\textbf {\bibinfo {volume}
  {100}},\ \bibinfo {pages} {025803} (\bibinfo {year} {2019})}\BibitemShut
  {NoStop}%
\bibitem [{\citenamefont {Pons}\ \emph {et~al.}(2013)\citenamefont {Pons},
  \citenamefont {Vigan{\`o}},\ and\ \citenamefont {Rea}}]{pons2013highly}%
  \BibitemOpen
  \bibfield  {author} {\bibinfo {author} {\bibfnamefont {J.~A.}\ \bibnamefont
  {Pons}}, \bibinfo {author} {\bibfnamefont {D.}~\bibnamefont {Vigan{\`o}}}, \
  and\ \bibinfo {author} {\bibfnamefont {N.}~\bibnamefont {Rea}},\ }\href@noop
  {} {\bibfield  {journal} {\bibinfo  {journal} {Nature Physics}\ }\textbf
  {\bibinfo {volume} {9}},\ \bibinfo {pages} {431} (\bibinfo {year}
  {2013})}\BibitemShut {NoStop}%
\bibitem [{\citenamefont {Horowitz}\ \emph {et~al.}(2015)\citenamefont
  {Horowitz}, \citenamefont {Berry}, \citenamefont {Briggs}, \citenamefont
  {Caplan}, \citenamefont {Cumming},\ and\ \citenamefont
  {Schneider}}]{PhysRevLett.114.031102}%
  \BibitemOpen
  \bibfield  {author} {\bibinfo {author} {\bibfnamefont {C.~J.}\ \bibnamefont
  {Horowitz}}, \bibinfo {author} {\bibfnamefont {D.~K.}\ \bibnamefont {Berry}},
  \bibinfo {author} {\bibfnamefont {C.~M.}\ \bibnamefont {Briggs}}, \bibinfo
  {author} {\bibfnamefont {M.~E.}\ \bibnamefont {Caplan}}, \bibinfo {author}
  {\bibfnamefont {A.}~\bibnamefont {Cumming}}, \ and\ \bibinfo {author}
  {\bibfnamefont {A.~S.}\ \bibnamefont {Schneider}},\ }\href {\doibase
  10.1103/PhysRevLett.114.031102} {\bibfield  {journal} {\bibinfo  {journal}
  {Phys. Rev. Lett.}\ }\textbf {\bibinfo {volume} {114}},\ \bibinfo {pages}
  {031102} (\bibinfo {year} {2015})}\BibitemShut {NoStop}%
\bibitem [{\citenamefont {Caplan}\ \emph {et~al.}(2018)\citenamefont {Caplan},
  \citenamefont {Schneider},\ and\ \citenamefont {Horowitz}}]{CaplanPRL}%
  \BibitemOpen
  \bibfield  {author} {\bibinfo {author} {\bibfnamefont {M.~E.}\ \bibnamefont
  {Caplan}}, \bibinfo {author} {\bibfnamefont {A.~S.}\ \bibnamefont
  {Schneider}}, \ and\ \bibinfo {author} {\bibfnamefont {C.~J.}\ \bibnamefont
  {Horowitz}},\ }\href {\doibase 10.1103/PhysRevLett.121.132701} {\bibfield
  {journal} {\bibinfo  {journal} {Phys. Rev. Lett.}\ }\textbf {\bibinfo
  {volume} {121}},\ \bibinfo {pages} {132701} (\bibinfo {year}
  {2018})}\BibitemShut {NoStop}%
\bibitem [{\citenamefont {Abbott}\ \emph {et~al.}(2019)\citenamefont {Abbott},
  \citenamefont {Abbott}, \citenamefont {Abbott}, \citenamefont {Abraham},
  \citenamefont {Acernese}, \citenamefont {Ackley}, \citenamefont {Adams},
  \citenamefont {Adhikari}, \citenamefont {Adya}, \citenamefont {Affeldt} \emph
  {et~al.}}]{abbott2019searches}%
  \BibitemOpen
  \bibfield  {author} {\bibinfo {author} {\bibfnamefont {B.}~\bibnamefont
  {Abbott}}, \bibinfo {author} {\bibfnamefont {R.}~\bibnamefont {Abbott}},
  \bibinfo {author} {\bibfnamefont {T.}~\bibnamefont {Abbott}}, \bibinfo
  {author} {\bibfnamefont {S.}~\bibnamefont {Abraham}}, \bibinfo {author}
  {\bibfnamefont {F.}~\bibnamefont {Acernese}}, \bibinfo {author}
  {\bibfnamefont {K.}~\bibnamefont {Ackley}}, \bibinfo {author} {\bibfnamefont
  {C.}~\bibnamefont {Adams}}, \bibinfo {author} {\bibfnamefont
  {R.}~\bibnamefont {Adhikari}}, \bibinfo {author} {\bibfnamefont
  {V.}~\bibnamefont {Adya}}, \bibinfo {author} {\bibfnamefont {C.}~\bibnamefont
  {Affeldt}},  \emph {et~al.},\ }\href@noop {} {\bibfield  {journal} {\bibinfo
  {journal} {The Astrophysical Journal}\ }\textbf {\bibinfo {volume} {879}},\
  \bibinfo {pages} {10} (\bibinfo {year} {2019})}\BibitemShut {NoStop}%
\bibitem [{\citenamefont {Pethick}(2019)}]{pethick2019dense}%
  \BibitemOpen
  \bibfield  {author} {\bibinfo {author} {\bibfnamefont {C.~J.}\ \bibnamefont
  {Pethick}},\ }\href@noop {} {\enquote {\bibinfo {title} {Dense matter and
  neutron stars: Some basic notions},}\ } (\bibinfo {year} {2019}),\ \Eprint
  {http://arxiv.org/abs/1912.11876} {arXiv:1912.11876 [nucl-th]} \BibitemShut
  {NoStop}%
\bibitem [{\citenamefont {Acevedo}\ \emph {et~al.}(2019)\citenamefont
  {Acevedo}, \citenamefont {Bramante}, \citenamefont {Leane},\ and\
  \citenamefont {Raj}}]{acevedo2019cooking}%
  \BibitemOpen
  \bibfield  {author} {\bibinfo {author} {\bibfnamefont {J.~F.}\ \bibnamefont
  {Acevedo}}, \bibinfo {author} {\bibfnamefont {J.}~\bibnamefont {Bramante}},
  \bibinfo {author} {\bibfnamefont {R.~K.}\ \bibnamefont {Leane}}, \ and\
  \bibinfo {author} {\bibfnamefont {N.}~\bibnamefont {Raj}},\ }\href@noop {}
  {\enquote {\bibinfo {title} {Cooking pasta with dark matter: Kinetic and
  annihilation heating of neutron star crusts},}\ } (\bibinfo {year} {2019}),\
  \Eprint {http://arxiv.org/abs/1911.06334} {arXiv:1911.06334 [hep-ph]}
  \BibitemShut {NoStop}%
\bibitem [{\citenamefont {Hanauske}\ \emph {et~al.}(2019)\citenamefont
  {Hanauske}, \citenamefont {Steinheimer}, \citenamefont {Motornenko},
  \citenamefont {Vovchenko}, \citenamefont {Bovard}, \citenamefont {Most},
  \citenamefont {Papenfort}, \citenamefont {Schramm},\ and\ \citenamefont
  {Stöcker}}]{Hanauske:2019qgs}%
  \BibitemOpen
  \bibfield  {author} {\bibinfo {author} {\bibfnamefont {M.}~\bibnamefont
  {Hanauske}}, \bibinfo {author} {\bibfnamefont {J.}~\bibnamefont
  {Steinheimer}}, \bibinfo {author} {\bibfnamefont {A.}~\bibnamefont
  {Motornenko}}, \bibinfo {author} {\bibfnamefont {V.}~\bibnamefont
  {Vovchenko}}, \bibinfo {author} {\bibfnamefont {L.}~\bibnamefont {Bovard}},
  \bibinfo {author} {\bibfnamefont {E.~R.}\ \bibnamefont {Most}}, \bibinfo
  {author} {\bibfnamefont {L.~J.}\ \bibnamefont {Papenfort}}, \bibinfo {author}
  {\bibfnamefont {S.}~\bibnamefont {Schramm}}, \ and\ \bibinfo {author}
  {\bibfnamefont {H.}~\bibnamefont {Stöcker}},\ }\bibfield  {booktitle} {\emph
  {\bibinfo {booktitle} {{Proceedings, The Modern Physics of Compact Stars and
  Relativistic Gravity 2017 (MPCS2017): Yerevan, Armenia, September 18-22,
  2017}}},\ }\href {\doibase 10.3390/particles2010004} {\bibfield  {journal}
  {\bibinfo  {journal} {Particles}\ }\textbf {\bibinfo {volume} {2}},\ \bibinfo
  {pages} {44} (\bibinfo {year} {2019})}\BibitemShut {NoStop}%
\bibitem [{\citenamefont {Schuetrumpf}\ \emph {et~al.}(2014)\citenamefont
  {Schuetrumpf}, \citenamefont {Iida}, \citenamefont {Maruhn},\ and\
  \citenamefont {Reinhard}}]{PhysRevC.90.055802}%
  \BibitemOpen
  \bibfield  {author} {\bibinfo {author} {\bibfnamefont {B.}~\bibnamefont
  {Schuetrumpf}}, \bibinfo {author} {\bibfnamefont {K.}~\bibnamefont {Iida}},
  \bibinfo {author} {\bibfnamefont {J.~A.}\ \bibnamefont {Maruhn}}, \ and\
  \bibinfo {author} {\bibfnamefont {P.-G.}\ \bibnamefont {Reinhard}},\ }\href
  {\doibase 10.1103/PhysRevC.90.055802} {\bibfield  {journal} {\bibinfo
  {journal} {Phys. Rev. C}\ }\textbf {\bibinfo {volume} {90}},\ \bibinfo
  {pages} {055802} (\bibinfo {year} {2014})}\BibitemShut {NoStop}%
\bibitem [{\citenamefont {Fattoyev}\ \emph {et~al.}(2017)\citenamefont
  {Fattoyev}, \citenamefont {Horowitz},\ and\ \citenamefont
  {Schuetrumpf}}]{PhysRevC.95.055804}%
  \BibitemOpen
  \bibfield  {author} {\bibinfo {author} {\bibfnamefont {F.~J.}\ \bibnamefont
  {Fattoyev}}, \bibinfo {author} {\bibfnamefont {C.~J.}\ \bibnamefont
  {Horowitz}}, \ and\ \bibinfo {author} {\bibfnamefont {B.}~\bibnamefont
  {Schuetrumpf}},\ }\href {\doibase 10.1103/PhysRevC.95.055804} {\bibfield
  {journal} {\bibinfo  {journal} {Phys. Rev. C}\ }\textbf {\bibinfo {volume}
  {95}},\ \bibinfo {pages} {055804} (\bibinfo {year} {2017})}\BibitemShut
  {NoStop}%
\bibitem [{\citenamefont {Schneider}\ \emph {et~al.}(2016)\citenamefont
  {Schneider}, \citenamefont {Berry}, \citenamefont {Caplan}, \citenamefont
  {Horowitz},\ and\ \citenamefont {Lin}}]{PhysRevC.93.065806}%
  \BibitemOpen
  \bibfield  {author} {\bibinfo {author} {\bibfnamefont {A.~S.}\ \bibnamefont
  {Schneider}}, \bibinfo {author} {\bibfnamefont {D.~K.}\ \bibnamefont
  {Berry}}, \bibinfo {author} {\bibfnamefont {M.~E.}\ \bibnamefont {Caplan}},
  \bibinfo {author} {\bibfnamefont {C.~J.}\ \bibnamefont {Horowitz}}, \ and\
  \bibinfo {author} {\bibfnamefont {Z.}~\bibnamefont {Lin}},\ }\href {\doibase
  10.1103/PhysRevC.93.065806} {\bibfield  {journal} {\bibinfo  {journal} {Phys.
  Rev. C}\ }\textbf {\bibinfo {volume} {93}},\ \bibinfo {pages} {065806}
  (\bibinfo {year} {2016})}\BibitemShut {NoStop}%
\bibitem [{\citenamefont {van Zoelen}\ and\ \citenamefont {ten
  Brinke}(2009)}]{van2009thin}%
  \BibitemOpen
  \bibfield  {author} {\bibinfo {author} {\bibfnamefont {W.}~\bibnamefont {van
  Zoelen}}\ and\ \bibinfo {author} {\bibfnamefont {G.}~\bibnamefont {ten
  Brinke}},\ }\href@noop {} {\bibfield  {journal} {\bibinfo  {journal} {Soft
  Matter}\ }\textbf {\bibinfo {volume} {5}},\ \bibinfo {pages} {1568} (\bibinfo
  {year} {2009})}\BibitemShut {NoStop}%
\bibitem [{\citenamefont {Berry}\ \emph {et~al.}(2016)\citenamefont {Berry},
  \citenamefont {Caplan}, \citenamefont {Horowitz}, \citenamefont {Huber},\
  and\ \citenamefont {Schneider}}]{PhysRevC.94.055801}%
  \BibitemOpen
  \bibfield  {author} {\bibinfo {author} {\bibfnamefont {D.~K.}\ \bibnamefont
  {Berry}}, \bibinfo {author} {\bibfnamefont {M.~E.}\ \bibnamefont {Caplan}},
  \bibinfo {author} {\bibfnamefont {C.~J.}\ \bibnamefont {Horowitz}}, \bibinfo
  {author} {\bibfnamefont {G.}~\bibnamefont {Huber}}, \ and\ \bibinfo {author}
  {\bibfnamefont {A.~S.}\ \bibnamefont {Schneider}},\ }\href {\doibase
  10.1103/PhysRevC.94.055801} {\bibfield  {journal} {\bibinfo  {journal} {Phys.
  Rev. C}\ }\textbf {\bibinfo {volume} {94}},\ \bibinfo {pages} {055801}
  (\bibinfo {year} {2016})}\BibitemShut {NoStop}%
\bibitem [{\citenamefont {Guven}\ \emph {et~al.}(2014)\citenamefont {Guven},
  \citenamefont {Huber},\ and\ \citenamefont
  {Valencia}}]{PhysRevLett.113.188101}%
  \BibitemOpen
  \bibfield  {author} {\bibinfo {author} {\bibfnamefont {J.}~\bibnamefont
  {Guven}}, \bibinfo {author} {\bibfnamefont {G.}~\bibnamefont {Huber}}, \ and\
  \bibinfo {author} {\bibfnamefont {D.~M.}\ \bibnamefont {Valencia}},\ }\href
  {\doibase 10.1103/PhysRevLett.113.188101} {\bibfield  {journal} {\bibinfo
  {journal} {Phys. Rev. Lett.}\ }\textbf {\bibinfo {volume} {113}},\ \bibinfo
  {pages} {188101} (\bibinfo {year} {2014})}\BibitemShut {NoStop}%
\bibitem [{\citenamefont {Schneider}\ \emph {et~al.}(2018)\citenamefont
  {Schneider}, \citenamefont {Caplan}, \citenamefont {Berry},\ and\
  \citenamefont {Horowitz}}]{Schneider:18}%
  \BibitemOpen
  \bibfield  {author} {\bibinfo {author} {\bibfnamefont {A.~S.}\ \bibnamefont
  {Schneider}}, \bibinfo {author} {\bibfnamefont {M.~E.}\ \bibnamefont
  {Caplan}}, \bibinfo {author} {\bibfnamefont {D.~K.}\ \bibnamefont {Berry}}, \
  and\ \bibinfo {author} {\bibfnamefont {C.~J.}\ \bibnamefont {Horowitz}},\
  }\href {\doibase 10.1103/PhysRevC.98.055801} {\bibfield  {journal} {\bibinfo
  {journal} {Phys. Rev. C}\ }\textbf {\bibinfo {volume} {98}},\ \bibinfo
  {pages} {055801} (\bibinfo {year} {2018})}\BibitemShut {NoStop}%
\bibitem [{\citenamefont {Schneider}\ \emph {et~al.}(2014)\citenamefont
  {Schneider}, \citenamefont {Berry}, \citenamefont {Briggs}, \citenamefont
  {Caplan},\ and\ \citenamefont {Horowitz}}]{PhysRevC.90.055805}%
  \BibitemOpen
  \bibfield  {author} {\bibinfo {author} {\bibfnamefont {A.~S.}\ \bibnamefont
  {Schneider}}, \bibinfo {author} {\bibfnamefont {D.~K.}\ \bibnamefont
  {Berry}}, \bibinfo {author} {\bibfnamefont {C.~M.}\ \bibnamefont {Briggs}},
  \bibinfo {author} {\bibfnamefont {M.~E.}\ \bibnamefont {Caplan}}, \ and\
  \bibinfo {author} {\bibfnamefont {C.~J.}\ \bibnamefont {Horowitz}},\ }\href
  {\doibase 10.1103/PhysRevC.90.055805} {\bibfield  {journal} {\bibinfo
  {journal} {Phys. Rev. C}\ }\textbf {\bibinfo {volume} {90}},\ \bibinfo
  {pages} {055805} (\bibinfo {year} {2014})}\BibitemShut {NoStop}%
\bibitem [{\citenamefont {Schneider}\ \emph {et~al.}(2013)\citenamefont
  {Schneider}, \citenamefont {Horowitz}, \citenamefont {Hughto},\ and\
  \citenamefont {Berry}}]{PhysRevC.88.065807}%
  \BibitemOpen
  \bibfield  {author} {\bibinfo {author} {\bibfnamefont {A.~S.}\ \bibnamefont
  {Schneider}}, \bibinfo {author} {\bibfnamefont {C.~J.}\ \bibnamefont
  {Horowitz}}, \bibinfo {author} {\bibfnamefont {J.}~\bibnamefont {Hughto}}, \
  and\ \bibinfo {author} {\bibfnamefont {D.~K.}\ \bibnamefont {Berry}},\ }\href
  {\doibase 10.1103/PhysRevC.88.065807} {\bibfield  {journal} {\bibinfo
  {journal} {Phys. Rev. C}\ }\textbf {\bibinfo {volume} {88}},\ \bibinfo
  {pages} {065807} (\bibinfo {year} {2013})}\BibitemShut {NoStop}%
\bibitem [{\citenamefont {{Horowitz}}\ \emph {et~al.}(2016)\citenamefont
  {{Horowitz}}, \citenamefont {{Berry}}, \citenamefont {{Caplan}},
  \citenamefont {{Fischer}}, \citenamefont {{Lin}}, \citenamefont {{Newton}},
  \citenamefont {{O'Connor}},\ and\ \citenamefont
  {{Roberts}}}]{Horowitz2016supernova}%
  \BibitemOpen
  \bibfield  {author} {\bibinfo {author} {\bibfnamefont {C.~J.}\ \bibnamefont
  {{Horowitz}}}, \bibinfo {author} {\bibfnamefont {D.~K.}\ \bibnamefont
  {{Berry}}}, \bibinfo {author} {\bibfnamefont {M.~E.}\ \bibnamefont
  {{Caplan}}}, \bibinfo {author} {\bibfnamefont {T.}~\bibnamefont {{Fischer}}},
  \bibinfo {author} {\bibfnamefont {Z.}~\bibnamefont {{Lin}}}, \bibinfo
  {author} {\bibfnamefont {W.~G.}\ \bibnamefont {{Newton}}}, \bibinfo {author}
  {\bibfnamefont {E.}~\bibnamefont {{O'Connor}}}, \ and\ \bibinfo {author}
  {\bibfnamefont {L.~F.}\ \bibnamefont {{Roberts}}},\ }\href@noop {} {\bibfield
   {journal} {\bibinfo  {journal} {arXiv e-prints}\ ,\ \bibinfo {eid}
  {arXiv:1611.10226}} (\bibinfo {year} {2016})},\ \Eprint
  {http://arxiv.org/abs/1611.10226} {arXiv:1611.10226 [astro-ph.HE]}
  \BibitemShut {NoStop}%
\bibitem [{\citenamefont {Grill}\ \emph {et~al.}(2012)\citenamefont {Grill},
  \citenamefont {Provid\^encia},\ and\ \citenamefont {Avancini}}]{Grill2012}%
  \BibitemOpen
  \bibfield  {author} {\bibinfo {author} {\bibfnamefont {F.}~\bibnamefont
  {Grill}}, \bibinfo {author} {\bibfnamefont {C.~m.~c.}\ \bibnamefont
  {Provid\^encia}}, \ and\ \bibinfo {author} {\bibfnamefont {S.~S.}\
  \bibnamefont {Avancini}},\ }\href {\doibase 10.1103/PhysRevC.85.055808}
  {\bibfield  {journal} {\bibinfo  {journal} {Phys. Rev. C}\ }\textbf {\bibinfo
  {volume} {85}},\ \bibinfo {pages} {055808} (\bibinfo {year}
  {2012})}\BibitemShut {NoStop}%
\bibitem [{\citenamefont {Horowitz}\ \emph {et~al.}(2004)\citenamefont
  {Horowitz}, \citenamefont {P\'erez-Garc\'{\i}a},\ and\ \citenamefont
  {Piekarewicz}}]{PhysRevC.69.045804}%
  \BibitemOpen
  \bibfield  {author} {\bibinfo {author} {\bibfnamefont {C.~J.}\ \bibnamefont
  {Horowitz}}, \bibinfo {author} {\bibfnamefont {M.~A.}\ \bibnamefont
  {P\'erez-Garc\'{\i}a}}, \ and\ \bibinfo {author} {\bibfnamefont
  {J.}~\bibnamefont {Piekarewicz}},\ }\href {\doibase
  10.1103/PhysRevC.69.045804} {\bibfield  {journal} {\bibinfo  {journal} {Phys.
  Rev. C}\ }\textbf {\bibinfo {volume} {69}},\ \bibinfo {pages} {045804}
  (\bibinfo {year} {2004})}\BibitemShut {NoStop}%
\bibitem [{\citenamefont {Caplan}\ \emph {et~al.}(2015)\citenamefont {Caplan},
  \citenamefont {Schneider}, \citenamefont {Horowitz},\ and\ \citenamefont
  {Berry}}]{PhysRevC.91.065802}%
  \BibitemOpen
  \bibfield  {author} {\bibinfo {author} {\bibfnamefont {M.~E.}\ \bibnamefont
  {Caplan}}, \bibinfo {author} {\bibfnamefont {A.~S.}\ \bibnamefont
  {Schneider}}, \bibinfo {author} {\bibfnamefont {C.~J.}\ \bibnamefont
  {Horowitz}}, \ and\ \bibinfo {author} {\bibfnamefont {D.~K.}\ \bibnamefont
  {Berry}},\ }\href {\doibase 10.1103/PhysRevC.91.065802} {\bibfield  {journal}
  {\bibinfo  {journal} {Phys. Rev. C}\ }\textbf {\bibinfo {volume} {91}},\
  \bibinfo {pages} {065802} (\bibinfo {year} {2015})}\BibitemShut {NoStop}%
\bibitem [{\citenamefont {Dorso}\ \emph {et~al.}(2018)\citenamefont {Dorso},
  \citenamefont {Frank},\ and\ \citenamefont {L{\'o}pez}}]{dorso2018phase}%
  \BibitemOpen
  \bibfield  {author} {\bibinfo {author} {\bibfnamefont {C.~O.}\ \bibnamefont
  {Dorso}}, \bibinfo {author} {\bibfnamefont {G.~A.}\ \bibnamefont {Frank}}, \
  and\ \bibinfo {author} {\bibfnamefont {J.~A.}\ \bibnamefont {L{\'o}pez}},\
  }\href@noop {} {\bibfield  {journal} {\bibinfo  {journal} {Nuclear Physics
  A}\ }\textbf {\bibinfo {volume} {978}},\ \bibinfo {pages} {35} (\bibinfo
  {year} {2018})}\BibitemShut {NoStop}%
\bibitem [{Note1()}]{Note1}%
  \BibitemOpen
  \bibinfo {note} {The minimum temperature is constrained by the model; at low
  $T$ the semi-classical model undergoes a phase transition to a solid, which
  we do not regard as physically relevant for nuclear physics, though this
  phase transition and the behavior of the model at low $T$ may be interesting
  if this model is used to study analagous systems, such as self-assembly in
  colloidal mixtures \cite
  {RevModPhys.89.041002,PhysRevC.94.055801}.}\BibitemShut {Stop}%
\bibitem [{SM()}]{SM}%
  \BibitemOpen
  \href@noop {} {}\bibinfo {note} {Animations available at: \\
  {\url{www.phy.ilstu.edu/~mcaplan/pasta-thermal/}}}\BibitemShut {NoStop}%
\bibitem [{\citenamefont {Caplan}\ and\ \citenamefont
  {Horowitz}(2017{\natexlab{b}})}]{RevModPhys.89.041002}%
  \BibitemOpen
  \bibfield  {author} {\bibinfo {author} {\bibfnamefont {M.~E.}\ \bibnamefont
  {Caplan}}\ and\ \bibinfo {author} {\bibfnamefont {C.~J.}\ \bibnamefont
  {Horowitz}},\ }\href {\doibase 10.1103/RevModPhys.89.041002} {\bibfield
  {journal} {\bibinfo  {journal} {Rev. Mod. Phys.}\ }\textbf {\bibinfo {volume}
  {89}},\ \bibinfo {pages} {041002} (\bibinfo {year}
  {2017}{\natexlab{b}})}\BibitemShut {NoStop}%
\bibitem [{\citenamefont {Lang}\ \emph {et~al.}(2001)\citenamefont {Lang},
  \citenamefont {Ohser},\ and\ \citenamefont {Hilfer}}]{lang2001analysis}%
  \BibitemOpen
  \bibfield  {author} {\bibinfo {author} {\bibfnamefont {C.}~\bibnamefont
  {Lang}}, \bibinfo {author} {\bibfnamefont {J.}~\bibnamefont {Ohser}}, \ and\
  \bibinfo {author} {\bibfnamefont {R.}~\bibnamefont {Hilfer}},\ }\href@noop {}
  {\bibfield  {journal} {\bibinfo  {journal} {Journal of microscopy}\ }\textbf
  {\bibinfo {volume} {203}},\ \bibinfo {pages} {303} (\bibinfo {year}
  {2001})}\BibitemShut {NoStop}%
\bibitem [{Note2()}]{Note2}%
  \BibitemOpen
  \bibinfo {note} {In our simulations we use $\lambda = 10$ fm for the
  proton-proton Coulomb screening. However, using $k_{TF}^{-1} = 11.5$ or
  $k_{TF}^{-1} = 10$ results in only a 2\% variation in $\Lambda
  _{ep}$.}\BibitemShut {Stop}%
\bibitem [{\citenamefont {Nandi}\ and\ \citenamefont
  {Schramm}(2018)}]{Nandi2018}%
  \BibitemOpen
  \bibfield  {author} {\bibinfo {author} {\bibfnamefont {R.}~\bibnamefont
  {Nandi}}\ and\ \bibinfo {author} {\bibfnamefont {S.}~\bibnamefont
  {Schramm}},\ }\href {\doibase 10.3847/1538-4357/aa9f12} {\bibfield  {journal}
  {\bibinfo  {journal} {The Astrophysical Journal}\ }\textbf {\bibinfo {volume}
  {852}},\ \bibinfo {pages} {135} (\bibinfo {year} {2018})}\BibitemShut
  {NoStop}%
\bibitem [{Note3()}]{Note3}%
  \BibitemOpen
  \bibinfo {note} {Although $\rho /\rho _0=0.3$ better matches the density we
  simulate in this work, $\rho /\rho _0=0.4$ is where the QMD model often finds
  the lasagna phase \cite {PhysRevC.66.012801, PhysRevC.68.035806}. Therefore,
  we look at both densities when making parallels between our results and those
  of Ref. \cite {Nandi2018}.}\BibitemShut {Stop}%
\bibitem [{\citenamefont {Maruyama}\ \emph {et~al.}(1998)\citenamefont
  {Maruyama}, \citenamefont {Niita}, \citenamefont {Oyamatsu}, \citenamefont
  {Maruyama}, \citenamefont {Chiba},\ and\ \citenamefont
  {Iwamoto}}]{PhysRevC.57.655}%
  \BibitemOpen
  \bibfield  {author} {\bibinfo {author} {\bibfnamefont {T.}~\bibnamefont
  {Maruyama}}, \bibinfo {author} {\bibfnamefont {K.}~\bibnamefont {Niita}},
  \bibinfo {author} {\bibfnamefont {K.}~\bibnamefont {Oyamatsu}}, \bibinfo
  {author} {\bibfnamefont {T.}~\bibnamefont {Maruyama}}, \bibinfo {author}
  {\bibfnamefont {S.}~\bibnamefont {Chiba}}, \ and\ \bibinfo {author}
  {\bibfnamefont {A.}~\bibnamefont {Iwamoto}},\ }\href {\doibase
  10.1103/PhysRevC.57.655} {\bibfield  {journal} {\bibinfo  {journal} {Phys.
  Rev. C}\ }\textbf {\bibinfo {volume} {57}},\ \bibinfo {pages} {655} (\bibinfo
  {year} {1998})}\BibitemShut {NoStop}%
\bibitem [{\citenamefont {Nandi}\ and\ \citenamefont
  {Schramm}(2016)}]{PhysRevC.94.025806}%
  \BibitemOpen
  \bibfield  {author} {\bibinfo {author} {\bibfnamefont {R.}~\bibnamefont
  {Nandi}}\ and\ \bibinfo {author} {\bibfnamefont {S.}~\bibnamefont
  {Schramm}},\ }\href {\doibase 10.1103/PhysRevC.94.025806} {\bibfield
  {journal} {\bibinfo  {journal} {Phys. Rev. C}\ }\textbf {\bibinfo {volume}
  {94}},\ \bibinfo {pages} {025806} (\bibinfo {year} {2016})}\BibitemShut
  {NoStop}%
\bibitem [{\citenamefont {Nandi}\ and\ \citenamefont
  {Schramm}(2017)}]{PhysRevC.95.065801}%
  \BibitemOpen
  \bibfield  {author} {\bibinfo {author} {\bibfnamefont {R.}~\bibnamefont
  {Nandi}}\ and\ \bibinfo {author} {\bibfnamefont {S.}~\bibnamefont
  {Schramm}},\ }\href {\doibase 10.1103/PhysRevC.95.065801} {\bibfield
  {journal} {\bibinfo  {journal} {Phys. Rev. C}\ }\textbf {\bibinfo {volume}
  {95}},\ \bibinfo {pages} {065801} (\bibinfo {year} {2017})}\BibitemShut
  {NoStop}%
\bibitem [{\citenamefont {Pethick}\ \emph {et~al.}(2020)\citenamefont
  {Pethick}, \citenamefont {Zhang},\ and\ \citenamefont
  {Kobyakov}}]{pethick2020elastic}%
  \BibitemOpen
  \bibfield  {author} {\bibinfo {author} {\bibfnamefont {C.}~\bibnamefont
  {Pethick}}, \bibinfo {author} {\bibfnamefont {Z.}~\bibnamefont {Zhang}}, \
  and\ \bibinfo {author} {\bibfnamefont {D.}~\bibnamefont {Kobyakov}},\
  }\href@noop {} {\bibfield  {journal} {\bibinfo  {journal} {arXiv preprint
  arXiv:2003.13430}\ } (\bibinfo {year} {2020})}\BibitemShut {NoStop}%
\bibitem [{\citenamefont {Reinhard}\ \emph {et~al.}(2006)\citenamefont
  {Reinhard}, \citenamefont {Bender}, \citenamefont {Nazarewicz},\ and\
  \citenamefont {Vertse}}]{PhysRevC.73.014309}%
  \BibitemOpen
  \bibfield  {author} {\bibinfo {author} {\bibfnamefont {P.-G.}\ \bibnamefont
  {Reinhard}}, \bibinfo {author} {\bibfnamefont {M.}~\bibnamefont {Bender}},
  \bibinfo {author} {\bibfnamefont {W.}~\bibnamefont {Nazarewicz}}, \ and\
  \bibinfo {author} {\bibfnamefont {T.}~\bibnamefont {Vertse}},\ }\href
  {\doibase 10.1103/PhysRevC.73.014309} {\bibfield  {journal} {\bibinfo
  {journal} {Phys. Rev. C}\ }\textbf {\bibinfo {volume} {73}},\ \bibinfo
  {pages} {014309} (\bibinfo {year} {2006})}\BibitemShut {NoStop}%
\bibitem [{\citenamefont {Nakazato}\ \emph {et~al.}(2011)\citenamefont
  {Nakazato}, \citenamefont {Iida},\ and\ \citenamefont
  {Oyamatsu}}]{PhysRevC.83.065811}%
  \BibitemOpen
  \bibfield  {author} {\bibinfo {author} {\bibfnamefont {K.}~\bibnamefont
  {Nakazato}}, \bibinfo {author} {\bibfnamefont {K.}~\bibnamefont {Iida}}, \
  and\ \bibinfo {author} {\bibfnamefont {K.}~\bibnamefont {Oyamatsu}},\ }\href
  {\doibase 10.1103/PhysRevC.83.065811} {\bibfield  {journal} {\bibinfo
  {journal} {Phys. Rev. C}\ }\textbf {\bibinfo {volume} {83}},\ \bibinfo
  {pages} {065811} (\bibinfo {year} {2011})}\BibitemShut {NoStop}%
\bibitem [{\citenamefont {Hill}\ and\ \citenamefont
  {Wheeler}(1953)}]{PhysRev.89.1102}%
  \BibitemOpen
  \bibfield  {author} {\bibinfo {author} {\bibfnamefont {D.~L.}\ \bibnamefont
  {Hill}}\ and\ \bibinfo {author} {\bibfnamefont {J.~A.}\ \bibnamefont
  {Wheeler}},\ }\href {\doibase 10.1103/PhysRev.89.1102} {\bibfield  {journal}
  {\bibinfo  {journal} {Phys. Rev.}\ }\textbf {\bibinfo {volume} {89}},\
  \bibinfo {pages} {1102} (\bibinfo {year} {1953})}\BibitemShut {NoStop}%
\bibitem [{\citenamefont {Rumyantsev}\ and\ \citenamefont
  {de~Pablo}(2020)}]{rumyantsev2020microphase}%
  \BibitemOpen
  \bibfield  {author} {\bibinfo {author} {\bibfnamefont {A.~M.}\ \bibnamefont
  {Rumyantsev}}\ and\ \bibinfo {author} {\bibfnamefont {J.~J.}\ \bibnamefont
  {de~Pablo}},\ }\href@noop {} {\bibfield  {journal} {\bibinfo  {journal}
  {Macromolecules}\ }\textbf {\bibinfo {volume} {53}},\ \bibinfo {pages} {1281}
  (\bibinfo {year} {2020})}\BibitemShut {NoStop}%
\bibitem [{\citenamefont {Watanabe}\ \emph {et~al.}(2002)\citenamefont
  {Watanabe}, \citenamefont {Sato}, \citenamefont {Yasuoka},\ and\
  \citenamefont {Ebisuzaki}}]{PhysRevC.66.012801}%
  \BibitemOpen
  \bibfield  {author} {\bibinfo {author} {\bibfnamefont {G.}~\bibnamefont
  {Watanabe}}, \bibinfo {author} {\bibfnamefont {K.}~\bibnamefont {Sato}},
  \bibinfo {author} {\bibfnamefont {K.}~\bibnamefont {Yasuoka}}, \ and\
  \bibinfo {author} {\bibfnamefont {T.}~\bibnamefont {Ebisuzaki}},\ }\href
  {\doibase 10.1103/PhysRevC.66.012801} {\bibfield  {journal} {\bibinfo
  {journal} {Phys. Rev. C}\ }\textbf {\bibinfo {volume} {66}},\ \bibinfo
  {pages} {012801} (\bibinfo {year} {2002})}\BibitemShut {NoStop}%
\bibitem [{\citenamefont {Watanabe}\ \emph {et~al.}(2003)\citenamefont
  {Watanabe}, \citenamefont {Sato}, \citenamefont {Yasuoka},\ and\
  \citenamefont {Ebisuzaki}}]{PhysRevC.68.035806}%
  \BibitemOpen
  \bibfield  {author} {\bibinfo {author} {\bibfnamefont {G.}~\bibnamefont
  {Watanabe}}, \bibinfo {author} {\bibfnamefont {K.}~\bibnamefont {Sato}},
  \bibinfo {author} {\bibfnamefont {K.}~\bibnamefont {Yasuoka}}, \ and\
  \bibinfo {author} {\bibfnamefont {T.}~\bibnamefont {Ebisuzaki}},\ }\href
  {\doibase 10.1103/PhysRevC.68.035806} {\bibfield  {journal} {\bibinfo
  {journal} {Phys. Rev. C}\ }\textbf {\bibinfo {volume} {68}},\ \bibinfo
  {pages} {035806} (\bibinfo {year} {2003})}\BibitemShut {NoStop}%
\end{thebibliography}%

\end{document}